\documentclass[aps,prb,superscriptaddress,amsmath,amssymb,reprint,]{revtex4-1}
\usepackage{amsmath}
\usepackage{txfonts}
\usepackage{amssymb}
\usepackage{braket}
\usepackage{natbib}
\usepackage{color,soul}       % use if color is used in text
\usepackage{textcomp}
\usepackage{graphicx}% Include figure files
\usepackage{dcolumn}% Align table columns on decimal point
\usepackage{bm}% bold math
\usepackage{multirow}
\usepackage{threeparttable}
\newcommand{\vQ}{v\left ( \mathbf{Q} \right )}

\newcommand{\Qe}{\left ( \mathbf{Q}, \epsilon \right )}

\newcommand{\kp}{\mathbf{k}}

\newcommand{\kpqp}{\mathbf{k'+q} }
\newcommand{\kpQp}{\mathbf{k'+Q} }

\newcommand{\kmQ}{\mathbf{k-Q}}
\newcommand{\kpp}{\mathbf{k'}}
\newcommand{\kppp}{\mathbf{k''}}
\newcommand{\kppqp}{\mathbf{k''+q}}
\newcommand{\kppQp}{\mathbf{k''+Q}}
\newcommand{\bmp}{\mathbf{m}}
\newcommand{\bn}{\mathbf{n}}
\newcommand{\br}{\mathbf{r} }
\newcommand{\brp}{\mathbf{r'}}

\newcommand{\go}{$G_oW_o$ }

\newcommand{\gowo}{$G_oW_o$ }
\newcommand{\gowop}{$G_oW_o$}
\newcommand{\goHFp}{$G_oW_o$@HF}
\newcommand{\goLDAp}{$G_oW_o$@LDA}
\newcommand{\goHF}{$G_oW_o$@HF }
\newcommand{\goLDA}{$G_oW_o$@LDA }
\newcommand{\goPWLDA}{$G_oW_o$@PW/LDA }

\newcommand{\goPBE}{$G_oW_o$@PBE }
\newcommand{\goPBEp}{$G_oW_o$@PBE}

\newcommand{\gw}{$GW$ }

\newcommand{\gwbse}{$GW$/BSE }

\newcommand{\gobse}{$G_oW_o$/BSE }
\newcommand{\kq}{\mathbf{k}+\mathbf{q}}
\newcommand{\kQ}{\mathbf{k}+\mathbf{Q}}
\newcommand{\kqQ}{\mathbf{k}+\mathbf{q}+\mathbf{Q}}
\newcommand{\kppq}{\mathbf{k''}+\mathbf{q''}}
\newcommand{\kppQ}{\mathbf{k''}+\mathbf{Q}}
\newcommand{\kppqQ}{\mathbf{k''}+\mathbf{q''}+\mathbf{Q}}
\newcommand{\mbQr}{e^{-i\mathbf{Q.r}}}

\newcommand{\bQr}{e^{i\mathbf{Q.r}}}
\newcommand{\bQrp}{e^{i\mathbf{Q.r'}}}
\newcommand{\bA}{\mathbf{A}}
\newcommand{\bB}{\mathbf{B}}
\newcommand{\bC}{\mathbf{C}}
\newcommand{\bQ}{\mathbf{Q}}
\newcommand{\bX}{\mathbf{X}}
\newcommand{\bY}{\mathbf{Y}}

\newcommand{\bkq}{\mathbf{k}+\mathbf{q}}

\newcommand{\bk}{\mathbf{k}}
\newcommand{\bq}{\mathbf{q}}
\newcommand{\bqpp}{\mathbf{q''}}
\newcommand{\bG}{\mathbf{G}}

\newcommand{\drdrp}{\int d\br d\brp}

\newcommand{\ewalda}{|\br - \brp - \bA|}
\newcommand{\ewaldb}{|\br - \brp - \bB|}

\newcommand{\GW}{$G^{}W^{}$ }

\begin{document}

\title{\goHF and BSE methods in periodic systems from Hartree-Fock theory: gaussian orbital and density fitting approach}

\date{\today} 

\author{Charles H. Patterson}

\affiliation{School of Physics, Trinity College Dublin, Dublin, D02 PN40, Ireland}

\begin{abstract}
The $GW$ method for calculating quasi-particle energies of solids commonly begin from a DFT Hamiltonian and Kohn-Sham orbitals
in a plane wave basis. Screening of the coulomb interaction is implemented using the inverse dielectric function in the 
random phase approximation (RPA). We present \go calculations which begin from the Hartree-Fock method in a basis of gaussian orbitals.
The screened coulomb interaction, $W$, is obtained using a $W$ = $v$ + $v\Pi v$ approach without invoking a plasmon pole approximation. 
The polarizability, $\Pi$, in $W$ is treated at the RPA level. RPA polarizabilities require solution of Bethe-Salpeter equations (BSE) 
for each unique \textbf{Q} point. A strategy for obtaining self-energies which are converged with respect to number of virtual states 
is employed in which \go yields the majority of the self-energy and the remaining part from high energy virtual levels is evaluated 
at second-order. The methods are evaluated by applying them to elemental semiconductors (C, Si) and oxides (MgO and anatase and 
rutile TiO$_2$). Common errors of HF theory applied to materials include overestimation of both the band gap and valence band widths. 
These are corrected in the approach employed here. Typically, the RPA screened interaction results in overestimation of band gaps 
while the \go self-energy band width renormalization yields band widths for diamond and Si which are in good agreement with experiment. 
HF calculations are performed in gaussian orbital basis sets and \go and BSE calculations are performed using density fitting 
with a coulomb metric. 
\end{abstract}

\email{Charles.Patterson@tcd.ie} 

\maketitle

\section{Introduction\label{introduction}}

\textit{Ab-initio} \textit{GW} and Bethe-Salpeter equation \cite{Bethe51} (BSE) methods have been applied to crystalline materials 
over the last 60 years \cite{Hedin65,Hybertsen85,Hybertsen86}. This has mostly been achieved using plane wave (PW) basis sets and 
DFT hamiltonians \cite{Hybertsen86,Hybertsen87} and PW/DFT implementations of \GW and BSE methods are available
in a number of codes, including BerkeleyGW\cite{Hybertsen86,Deslippe12}, Yambo \cite{Marini09,Sangalli19}, VASP \cite{Shishkin07,Sander15} 
and Abinit \cite{Bruneval06,Gonze20}. Here we report \gobse calculations for tetrahedral semiconductors (Si and C) and wide gapped 
oxides (MgO and TiO$_2$) in a gaussian orbital (GO) basis using a density fitting approach using the Exciton code for both single-particle 
and \gobse calculations \cite{Patterson10,Patterson19,Patterson20a}. In contrast to most previous applications of the \gobse method in solids,
we use Hartree-Fock (HF) wavefunctions and single-particle energies in the unperturbed hamiltonian. The HF method is well known
to overestimate band gaps and valence band widths in solids and we show that the \go self-energies that we employ
are capable of renormalizing the bandwidths of Si and diamond, to yield agreement with experimental bandwidths,
as well as yielding \go band structures in reasonable agreement band structures from PW \go calculations.

The $\bQ$-dependent screened interaction, $W(\bQ$), in both \gw and BSE calculations in this work is obtained from the interacting 
polarizabilty in the random phase approximations (RPA). This approach has the advantage that it does not require a plasmon pole approximation 
to the inverse dielectric function, however, it does require diagonalization of large RPA hamiltonians at all unique $\bQ$ points 
in the Brillouin zone (BZ) while generating the BSE hamiltonian. This approach can also be applied to calculation of exciton dispersion.

Density fitting of products of wavefunction orbitals is a long established technique in finite 
\cite{Whitten73,Dunlap79,Mintmire82,Reine08} and periodic systems
%\cite{Whitten73,Dunlap79,Mintmire82a,Dunlap00,Jung05,Reine08,Koster09,Pedersen09,MejiaRodriguez14,Wirz17} %Geudtner12 
and periodic systems \cite{Maschio07,Usvyat07,Milko07,Varga08,Burow09,Katouda10,Lorenz12,DelBen13,Francini14,Sun17,Wang20,Patterson20a}.
%\cite{Mintmire82,Rohlfing95,Maschio07,Usvyat07,Milko07,Maschio08,Varga08,Burow09,Dunlap10,Katouda10,Maschio11,Lorenz12,DelBen13,Francini14,Sun17a,Wang20}.
The GO/density fitting approach has the advantage of a much reduced number of integrals required over approaches 
that do not factorize them. It lends itself to calculations involving other interacting particles such as positrons \cite{Hofierka22}, 
since the products above can be calculated separately for electron or positron wavefunctions. It may be possible to 
treat systems with large unit cells and open volumes such as polymers, amorphous and crystalline organic materials or
metal organic frameworks. Gaussian basis sets in quantum chemistry are commonly generated using variational
principles for the ground state, but this leaves open the question of their suitablility for conduction bands of solids.
We address this question here by calculating the free electron band structures of the materials chosen for study.

The present work follows earlier work in which density fitting was applied to time-dependent Hartree-Fock (TDHF) calculations
\cite{Patterson20a} where screening of the electron-hole attraction term in the TDHF Hamiltonain was implemented using a simple
scaling factor. It employs coulomb-weighted density fitting in which the metric is the coulomb potential, rather than the 
overlap of the auxiliary basis set. Using DFT wavefunctions and single-particle energies in the unperturbed hamiltonian, 
Rohlfing and coworkers showed that a GO basis could reproduce planewave/DFT quasiparticle (QP) energy corrections 
in tetrahedral semiconductors \cite{Rohlfing93,Rohlfing95} and oxides \cite{Rohlfing98a} as well as excitonic optical spectra 
in wide and narrow gapped materials \cite{Rohlfing00}. More recently, Zhu and Chan \cite{Zhu21} reported \goPBE calculations of 
valence and core level ionization potentials \cite{Zhu21}, Zgid and coworkers reported finite temperature \cite{Yeh22} and 
relativistic \cite{Abraham24} self-consistent $scGW$ calculations and Garc\'{\i}a-Bl\'azquez and Palacios 
\cite{GarciaBlazquez25} reported BSE calculations, all of which were in a GO basis with density fitting. Correlated electron methods 
such as coupled cluster singles and doubles (CCSD) for periodic systems \cite{Ye24} have also been implemented in a GO basis. 

The remainder of this paper is organized as follows: the BSE for finite wavevector excitations is introduced using a linear response
formalism, followed by the bare and interacting polarizabilities and second order and \go self-energies. The following section
describes how these are implemented in a GO basis using density fitting. Results of \go self-energy and 
BSE calculations are reported in the following sections for diamond and Si, MgO and anatase and rutile TiO$_2$. The final section
provides discussion and further analysis of these results and conclusions regarding the GO/density fitting approach
to excitations in gapped materials.

\section{Theory}

The BSE formalism is commonly introduced beginning from the two-body Green's function \cite{Strinati88,Hybertsen86,Sander15},
which leads to the well-known form,

\begin{equation}
\Pi\Qe = \Pi_0\Qe + \Pi_0\Qe K\Qe \Pi\Qe,
\label{eqn1}
\end{equation}

\noindent
where $\Pi\Qe$ and $\Pi_0\Qe$ are, respectively, the interacting and non-interacting polarizability and $\bQ$ and $\epsilon$ are 
wave vector and energy. The choice of interaction kernel, $K\Qe$, determines the level of approximation for $\Pi\Qe$.
In an RPA calculation, $K(\bQ)$ contains the static, bare 
coulomb interaction, $v(\bQ)$, in ring diagrams; in a BSE calculation $K\Qe$ contains $v(\bQ)$ in ring diagrams as well as the 
energy-dependent screened electron-hole interaction, $W\Qe$.

\begin{figure}[htp]
\includegraphics[width=4cm,angle=0]{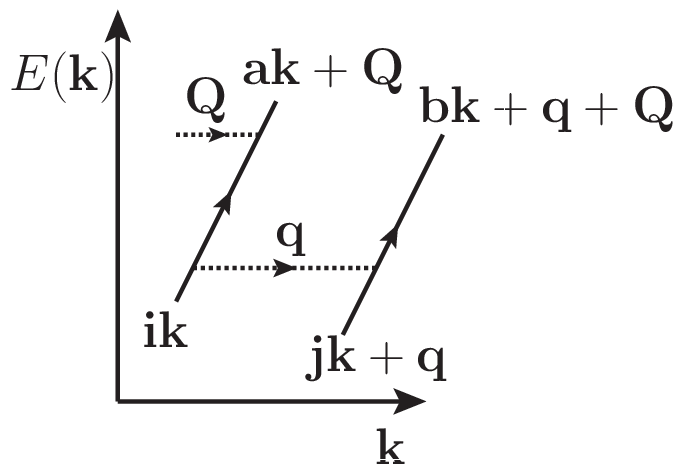}
\caption{Band energy versus wave vector diagram for indirect transitions with change in wave vector $\mathbf{Q}$ 
between states $\psi_{i\mathbf{k}}$ and $\psi_{a\mathbf{k+Q}}$ and states $\psi_{j\mathbf{k+q}}$ and $\psi_{b\mathbf{k+q+Q}}$. These
electron-hole pairs differ in wave vector $\mathbf{q}$.}
\label{fig1}
\end{figure}

Fig. \ref{fig1} shows two indirect transitions in which electron and hole wavevectors for each electron-hole pair differ 
by $\bQ$ while the wavevector connecting each electron-hole pair is $\bq$. This is the meaning of $\bq$ and $\bQ$ in this work.

In a conventional \gobse calculation of the optical properties of a material, $\Pi\Qe$ is calculated for $\bQ\to\mathbf{0}$. 
Energy denominators in $\Pi_0\Qe$ are single particle Kohn-Sham eigenvalues with \go self-energy corrections. 
The screened interaction, $W_0(\bQ,\epsilon)$, is calculated using the RPA inverse dielectric function, 

\begin{equation}
W_0(\bQ,\epsilon) = \epsilon^{-1,RPA}(\bQ,\epsilon) v(\bQ)
\label{eqn2}
\end{equation}

\noindent
where

\begin{equation}
\epsilon^{RPA}(\bQ,\epsilon) = 1 - v(\bQ) \Pi_0(\bQ,\epsilon)
\label{eqn3}
\end{equation}

\noindent
The frequency dependent inverse dielectric function in Eq. \ref{eqn2} is needed for the \go self-energy, which is the convolution of
$G_0$ and $W_0$. Inversion of the dielectric matrix as a function of frequency can be avoided using a plasmon pole
approximation, whereby the dielectric matrix is inverted at zero frequency and a frequency dependence $ansatz$ is introduced,
for example, by several pole frequencies whose strengths and positions are fitted using sum rules \cite{Hybertsen87}.

We adopt an approach in which the screened interaction is given by the equivalent form,

\begin{equation}
W_0\Qe = \vQ + \vQ \Pi\Qe \vQ,
\label{eqn4}
\end{equation}

\noindent
where $\Pi$ is the RPA dressed polarizability obtained by solving RPA equations at unique, finite wave vectors, $\bQ$. 
This screened interaction is used in the final BSE that is solved and used to obtain the imaginary part
of the optical dielectric function at $\bQ\to\mathbf{0}$. This approach requires no plasmon pole approximation since the 
frequency dependence of $W_0(\bQ,\epsilon)$ is obtained exactly (within the limitations of the RPA and basis sets used) when the 
self-energy is constructed from RPA eigenvectors and eigenvalues. 
A similar approach was applied recently to ionization and excitation energies in medium sized molecules \cite{Patterson24,Waide24}.

\subsection{Linear response equation of motion\label{linear}}

A linear response approach which emphasizes the connection of the BSE to linear response theory has been reviewed by 
Dreuw and Head-Gordon \cite{Dreuw05}. It yields the BSE equations for finite wave vector $\bQ$ excitations in a transparent manner,
beginning from the equation of motion for the density matrix,

\begin{equation}
i \frac{\partial \hat{\mathbf{P}}}{\partial t} = \left[ \hat{\mathbf{F}},\hat{\mathbf{P}} \right].
\label{eqn5}
\end{equation}

\noindent
This is solved in the basis of eigenfunctions of the Fock operator \{$i,a$\}. The Fock operator, $\hat{\mathbf{F}}$, and density 
matrix, $\hat{\mathbf{P}}$, are split into zeroth $(0)$ and first order $(1)$ parts, so that an element of the density matrix 
is given by,

\begin{equation}
P^{}_{pq} = P^{(0)}_{pq} + P^{(1)}_{pq}.
\label{eqn6}
\end{equation}

\noindent
The zeroth order Fock matrix, representing the ground state, is diagonal with elements,

\begin{equation}
F^{(0)}_{pq} = \epsilon_p \delta_{pq}.
\label{eqn7}
\end{equation}

\noindent
where $\epsilon_p$ are occupied and virtual state HF eigenvalues, $\epsilon_i$ and $\epsilon_a$. The zeroth order density matrix 
in the \{$i,a$\} basis is diagonal with unit values for states which are occupied in the ground state,

\begin{equation}
P^{(0)}_{pq} = \delta_{pq}\theta(\epsilon_{k_F} - \epsilon_p),
\label{eqn8}
\end{equation}

\noindent
and zero otherwise. $\epsilon_{k_F}$ is the Fermi energy. The electron-electron part of the zeroth order Fock matrix is,

\begin{widetext}
\begin{equation}
F^{(0)}_{p\kQ q\bk} = \left[ 2(\psi^*_{p\kQ} \psi^{}_{q\bk} |V|\psi^*_{r\kq} \psi^{}_{s\kqQ}) - \\
                              (\psi^*_{p\kQ} \psi^{}_{s\kqQ} |W|\psi^*_{r\kq} \psi^{}_{q\bk}) \right ]  P_{s\kqQ r\kq}, 
\label{eqn9}
\end{equation}
\end{widetext}

\noindent
The static, screened interaction, $W$, in the direct term in $F^{(0)}$ is introduced here on the basis that 
screening of these terms is found in derivation of the BSE beginning from the two-body Green's function and that this
is usually approximated by the static, screened interaction which enables solution of the resulting equations as a generalized
eigenvalue problem \cite{Hybertsen86}. Two-electron integrals in the Fock matrix are given in chemists' notation which lends 
itself to the density fitting approach outlined below. The first order Fock matrix is given by,

\begin{equation}
F^{(1)}_{ai} = \frac{\partial F^{(0)}_{ai}}{\partial P_{jb}} P^{(1)}_{jb} + \frac{\partial F^{(0)}_{ai}}{\partial P_{bj}} P^{(1)}_{bj},
\label{eqn10}
\end{equation}

\noindent 
where the first order density matrix is,

\begin{equation}
P^{(1)}_{bj}(t) = \frac{1}{2} \left[ X_{bj} e^{-i\epsilon t} + Y^*_{jb} e^{+i\epsilon t} \right ].
\label{eqn11}
\end{equation}

\noindent
$X$ and $Y$ are constant amplitudes to be determined from the equation of motion and $P^{(1)}_{bj}(t) = P^{*(1)}_{jb}(t)$. 
The system is assumed to be initially in its ground state, so that $P^{(0)}_{ii}$ = 1 and $P^{(0)}_{aa}$ is zero, 
and to remain predominantly in its ground state, so that $P^{(1)}_{aa}$ remains negligible. Off-diagonal elements of the commutators 
at first order are given by,

\begin{equation}
\left( F^{(0)}P^{(1)} - P^{(1)}F^{(0)} \right)_{ai} = \epsilon_a P^{(1)}_{ai} - P^{(1)}_{ai} \epsilon_i,
\label{eqn12}
\end{equation}

and

\begin{equation}
\left( F^{(1)}P^{(0)} - P^{(0)}F^{(1)} \right)_{ai} = F^{(1)}_{ai}P^{(0)}_{ii} - P^{(0)}_{aa}F^{(1)}_{ai}.
\label{eqn13}
\end{equation}

We now consider in more detail the equation of motion for excitations at wave vector $\bQ$ in a crystal. 
%The first order density at wave vector $\bQ$ is,
%\begin{widetext}
%\begin{equation}
%\rho^{(1)}(\br,t) = \frac{1}{2} \sum_{b,j,\bk,\bq} \left[ X^{}_{b\kqQ^{}j^*\kq} \psi^{}_{b\kqQ}(\br)\psi^*_{j\kq}(\br) e^{-i\epsilon t} + \
%                                                          Y^*_{j\kq^{}b^*\kqQ}  \psi^{}_{j\kq}(\br)\psi^*_{b\kqQ}(\br) e^{+i\epsilon t} 
%\right ] + c.c.
%\label{eqn13}
%\end{equation}
%\noindent
%$\bk$ and $\kq$ are a pair of wave vectors in the first BZ and $\bQ$ is fixed. 
A complex conjugate pair of off-diagonal, first order density matrix elements is,

\begin{equation}
P_{j^{}\kq b^*\kqQ}^{(1)}(t) = \frac{1}{2} \left[ X^{}_{j^{}\kq b^*\kqQ} e^{-i\epsilon t} + Y^*_{j^{}\kq b^*\kqQ} e^{+i\epsilon t} \right ] 
\label{eqn14}
\end{equation}

\noindent 
and

\begin{equation}
P_{b\kqQ^{}j^*\kq}^{(1)}(t) = \frac{1}{2} \left[ X^*_{b\kqQ^{}j^*\kq} e^{+i\epsilon t} + Y^{}_{b^{}\kqQ j^*\kq} e^{-i\epsilon t} \right ] 
\label{eqn15}
\end{equation}

Inserting the density matrix into Eq. \ref{eqn12} yields,

\begin{widetext}

\begin{equation}
F^{(0)}_{a\kQ,a\kQ}P^{(1)}_{a\kQ,i\bk}(t) - P^{(1)}_{a\kQ,i\bk}(t) F^{(0)}_{i\bk,i\bk} = \
(\epsilon_{a\kQ}  - \epsilon_{i\bk}) P^{(1)}_{a\kQ,i\bk}(t),
\label{eqn16}
\end{equation}

Eq. \ref{eqn13} becomes,

\begin{eqnarray}
\label{eqn17}
F^{(1)}_{a\kQ,i\mathbf{k}}P^{(0)}_{i\bk,i\bk} = \
  \left[ (\psi^*_{a\kQ}\psi^{}_{i\mathbf{k}}|V|\psi^*_{j\kq}\psi^{}_{b\kqQ}) - \
         (\psi^*_{a\kQ}\psi^{}_{b\kqQ}|W|\psi^*_{j\kq}\psi^{}_{i\mathbf{k}}) \right] P^{(1)}_{b\kqQ^{}j^*\kq} \\
\nonumber
+ \left[ (\psi^*_{a\kQ}\psi^{}_{i\mathbf{k}}|V|\psi^*_{b\kq}\psi^{}_{j\kqQ}) - \
         (\psi^*_{a\kQ}\psi^{}_{j\kqQ}|W|\psi^*_{b\kq}\psi^{}_{i\mathbf{k}}) \right] P^{(1)}_{j^{}\kqQ b^*\kq}.
\end{eqnarray}

Inserting Eq. \ref{eqn16} and \ref{eqn17} into Eq. \ref{eqn5} and equating coefficients of $e^{-i\epsilon t}$ with space variation 
$e^{i\bQ.\br}$ yields,

\begin{eqnarray}
(\epsilon_{a\kQ}  - \epsilon_{i\bk} - \epsilon) X_{i^{}\bk,a^*\kQ} + 
  \left[ (\psi^*_{a\kQ}\psi^{}_{i\mathbf{k}}|V|\psi^*_{j\kq}\psi^{}_{b\kqQ}) - \
         (\psi^*_{a\kQ}\psi^{}_{b\kqQ}|W|\psi^*_{j\kq}\psi^{}_{i\mathbf{k}}) \right] X_{j\kq^{}b^*\kqQ} \\
\nonumber 
+ 
\left[ (\psi^*_{a\kQ}\psi^{}_{i\mathbf{k}}|V|\psi^*_{b\kq}\psi^{}_{j\kqQ}) - \
       (\psi^*_{a\kQ}\psi^{}_{j\kqQ}|W|\psi^*_{b\kq}\psi^{}_{i\mathbf{k}}) \right] Y_{j^*\kqQ b^{}\kq}.
\label{eqn18}
\end{eqnarray}

A similar analysis of the $i\bk,a\kQ$ element of the density matrix in Eq. \ref{eqn5} yields,

\begin{eqnarray}
(\epsilon_{a\kQ}  - \epsilon_{i\bk} + \epsilon) Y_{a\kQ,i\bk} + 
  \left[ (\psi^*_{i\kQ}\psi^{}_{a\mathbf{k}}|V|\psi^*_{j\kq}\psi^{}_{b\kqQ}) - \
         (\psi^*_{i\kQ}\psi^{}_{b\kqQ}|W|\psi^*_{j\kq}\psi^{}_{a\mathbf{k}}) \right] X_{j\kq^{}b^*\kqQ} \\
\nonumber 
+ 
\left[ (\psi^*_{i\kQ}\psi^{}_{a\mathbf{k}}|V|\psi^*_{b\kq}\psi^{}_{j\kqQ}) - \
       (\psi^*_{i\kQ}\psi^{}_{j\kqQ}|W|\psi^*_{b\kq}\psi^{}_{a\mathbf{k}}) \right] Y_{j^*\kqQ b^{}\kq}.
\label{eqn19}
\end{eqnarray}

\begin{table}[h!]
\caption{Two-electron matrix elements in $A$ and $B$ blocks of the BSE matrix (Eq. \ref{eqn21}) used for calculation of the RPA-TDA
polarizability, $\Pi$, and for BSE-TDA excited state calculations. Factors of 2 appear from spin summation. The diagonals of the $A$ and 
$A^*$ blocks also contain eigenvalue differences, $(\epsilon_{a\kQ} - \epsilon_{i\bk}) \delta_{a\kQ b\kQ} \delta_{i\bk j\bk} $.
For RPA-TDA polarizability calculations these are HF single-particle eigenvalues and for BSE-TDA they are HF eigenvalues with
\goHF self-energy corrections.
}
\begin{ruledtabular}
\begin{tabular}{lcccccccccc} 
 Method     &     $A_{ai,bj}$     &  $A_{ai,bj}$        &    $B_{ai,bj}$      &   $B_{ai,bj}$      \\  
\hline     
 TDHF       & 2$(\psi^*_{a\kQ} \psi^{}_{i\bk} |V|\psi^*_{j\kq} \psi^{}_{b\kqQ}) $ 
            & -$(\psi^*_{a\kQ} \psi^{}_{b\kqQ}|V|\psi^*_{j\kq} \psi^{}_{i\bk})  $ 
            & 2$(\psi^*_{a\kQ} \psi^{}_{i\bk} |V|\psi^*_{b\kq} \psi^{}_{j\kqQ}) $ 
            & -$(\psi^*_{a\kQ} \psi^{}_{j\kqQ}|V|\psi^*_{b\kq} \psi^{}_{i\bk})  $ \\
 BSE        & 2$(\psi^*_{a\kQ} \psi^{}_{i\bk} |V|\psi^*_{j\kq} \psi^{}_{b\kqQ}) $ 
            & -$(\psi^*_{a\kQ} \psi^{}_{b\kqQ}|W|\psi^*_{j\kq} \psi^{}_{i\bk})  $ 
            & 2$(\psi^*_{a\kQ} \psi^{}_{i\bk} |V|\psi^*_{b\kq} \psi^{}_{j\kqQ}) $ 
            & -$(\psi^*_{a\kQ} \psi^{}_{j\kqQ}|W|\psi^*_{b\kq} \psi^{}_{i\bk})  $ \\
 RPA        & 2$(\psi^*_{a\kQ} \psi^{}_{i\bk} |V|\psi^*_{j\kq} \psi^{}_{b\kqQ}) $ 
            &
            & 2$(\psi^*_{a\kQ} \psi^{}_{i\bk} |V|\psi^*_{b\kq} \psi^{}_{j\kqQ}) $ 
            &                                                                   \\
 TDHF-TDA   & 2$(\psi^*_{a\kQ} \psi^{}_{i\bk} |V|\psi^*_{j\kq} \psi^{}_{b\kqQ}) $ 
            & -$(\psi^*_{a\kQ} \psi^{}_{b\kqQ}|V|\psi^*_{j\kq} \psi^{}_{i\bk})  $ 
            &
            &                                                                   \\
 BSE-TDA    & 2$(\psi^*_{a\kQ} \psi^{}_{i\bk} |V|\psi^*_{j\kq} \psi^{}_{b\kqQ}) $ 
            & -$(\psi^*_{a\kQ} \psi^{}_{b\kqQ}|W|\psi^*_{j\kq} \psi^{}_{i\bk})  $ 
            &
            &                                                                   \\
 RPA-TDA    & 2$(\psi^*_{a\kQ} \psi^{}_{i\bk} |V|\psi^*_{j\kq} \psi^{}_{b\kqQ}) $ 
            &
            &
            &                                                                   \\
\end{tabular}
\end{ruledtabular}
\label{tab1}
\end{table}

\end{widetext}

\noindent
Eq. \ref{eqn18} and \ref{eqn19} and their complex conjugates can be written in the form,

\begin{eqnarray}
\label{eqn20}
\bA \bX + \bB \bY &=& +\mathbf{\Omega} \bX\\
\nonumber 
\bB^* \bX + \bA^* \bY &=& -\mathbf{\Omega} \bY\\
\nonumber 
\bA^* \bX^* + \bB^* \bY^* &=& -\mathbf{\Omega} \bX^*\\
\nonumber 
\bB \bX^* + \bA \bY^* &=& +\mathbf{\Omega} \bY^*
\end{eqnarray}

\begin{figure}[htp]
\includegraphics[width=8.5cm,angle=0]{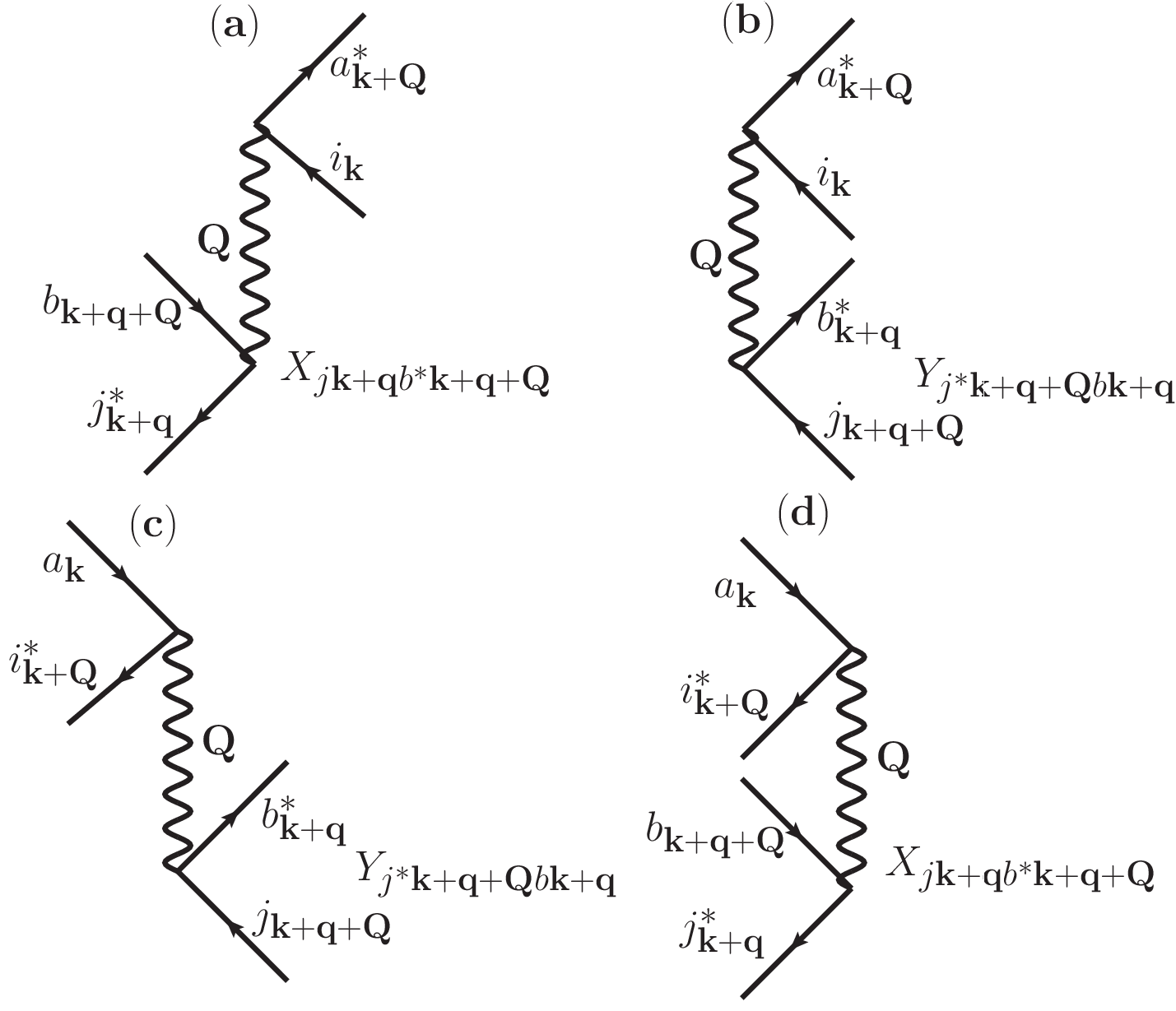}
\caption{Feynman diagrams which contribute to the exchange parts of Fock matrix elements $\mathbf{F}_{a\kQ,i\mathbf{k}}$ and 
$\mathbf{F}_{i\kQ,a\mathbf{k}}$. (a) $\mathbf{A X}$, (b) $\mathbf{B Y}$ (c) $\mathbf{A}^* \mathbf{Y}$ and (d) $\mathbf{B}^* \mathbf{X}$.
Momentum transfer is indicated by $\mathbf{Q}$.}
\label{fig2}
\end{figure}

\begin{figure}[htp]
\includegraphics[width=8.5cm,angle=0]{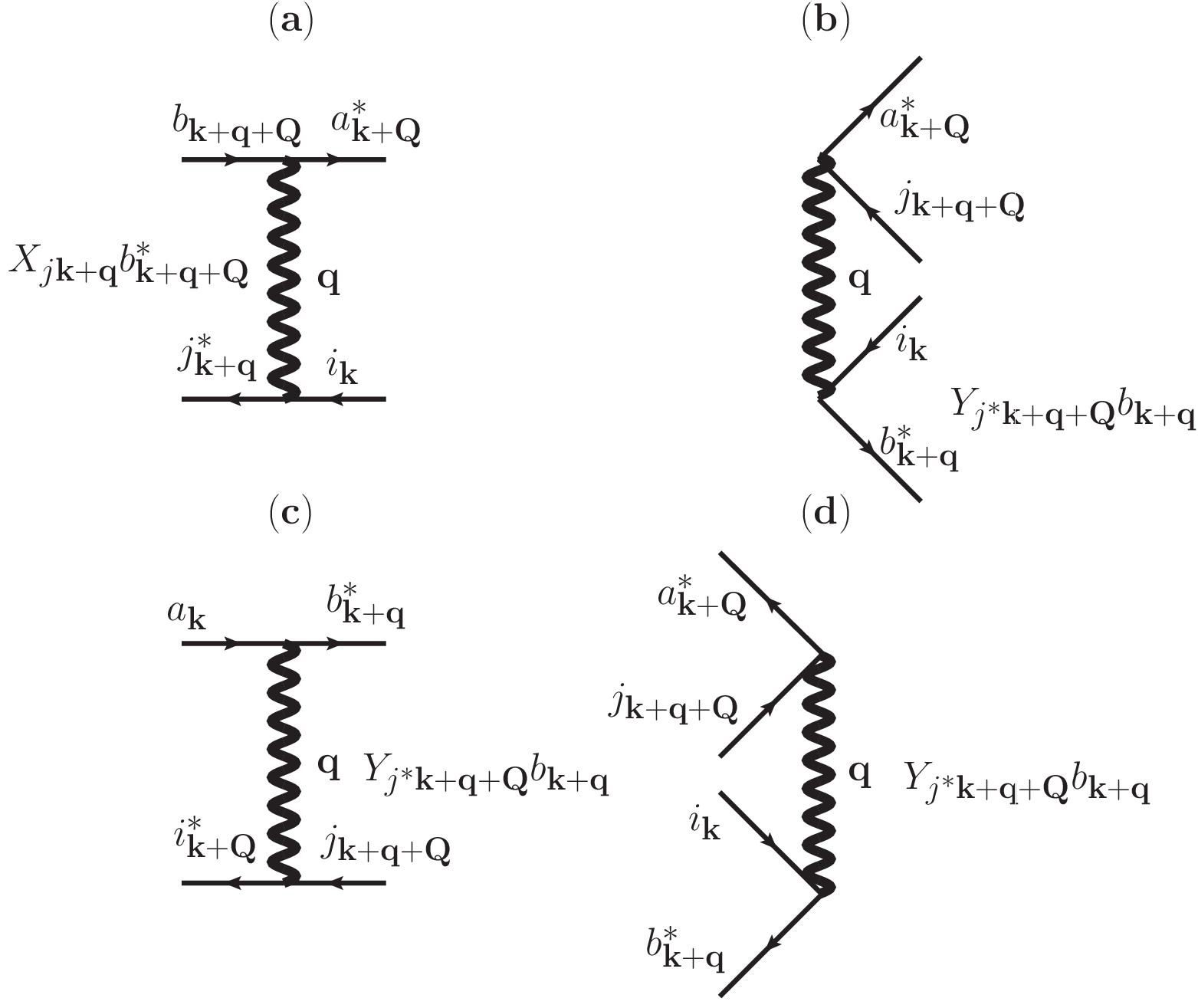}
\caption{Feynman diagrams which contribute to the direct parts of Fock matrix elements $\mathbf{F}_{a\kQ,i\mathbf{k}}$ and 
$\mathbf{F}_{i\kQ,a\mathbf{k}}$. (a) $\mathbf{A X}$, (b) $\mathbf{B Y}$ (c) $\mathbf{A}^* \mathbf{Y}$ and (d) $\mathbf{B}^* \mathbf{X}$.
Momentum transfer is indicated by $\mathbf{q}$.}
\label{fig3}
\end{figure}

\noindent
The correspondence between two-electron matrix elements in Eq. \ref{eqn18} and \ref{eqn19} and the $\bA$ and $\bB$ matrix blocks in
Eq. \ref{eqn20} is given in Table \ref{tab1} as well as further Tamm-Dancoff (TDA), RPA and TDHF approximations.
Feynman diagram representations of the matrix elements in Eq. \ref{eqn18} and \ref{eqn19} are shown in Figs. \ref{fig2} and \ref{fig3}.
The system of coupled equations in Eq. \ref{eqn20} can be represented as the BSE,

\begin{equation}
\left( \begin{array}{cc}
\mathbf{A} & \mathbf{B} \\
\mathbf{B}^* & \mathbf{A}^* \end{array} \right)
\left( \begin{array}{cc}
\mathbf{X} & \mathbf{Y^*} \\ \mathbf{Y} & \mathbf{X^*} \end{array} \right) = 
\left( \begin{array}{cc}
\mathbf{\Omega} & \mathbf{0} \\ \mathbf{0} & -\mathbf{\Omega} \end{array} \right) 
\left( \begin{array}{cc}
\mathbf{X} & \mathbf{Y^*} \\ \mathbf{Y} & \mathbf{X^*} \end{array} \right) 
\label{eqn21}
\end{equation}

\noindent
%Elements of the $\bA$ and $\bB$ sub-blocks of the BSE matrix are given in Table \ref{tab1}.
Eq. \ref{eqn21} is a generalized eigenvalue problem in which eigenvalues occur in positive and negative pairs, which we denote
as $\Omega^S_\pm$, with $\Omega^S_+$ denoting the positive eigenvalue of the pair.
Components of the $S^{th}$ eigenvector at wave vector $\bQ$ are denoted, $\mathbf{X}_{i\bk a\kQ}$ and $\mathbf{Y}_{\bk ia\kQ}$,
(Eq. \ref{eqn21}). The corresponding eigenvalue is denoted, $\Omega^S(\bQ)$. 
%The combination, 

%\begin{equation}
%\left[ \mathbf{X}^{}_{i\bk a\kQ}e^{-i\epsilon t} + \mathbf{Y}^\dagger_{i\bk a\kQ} e^{+i\epsilon t} \right ]
%\psi^{}_{i\bk}(\mathbf{x})\psi^*_{a\kQ}(\mathbf{x}),
%\label{eqn23}
%\end{equation}

%\noindent
%is a particle-hole creation (excitation) amplitude and the corresponding de-excitation amplitude is,

%\begin{equation}
%\left[ \mathbf{X}^\dagger_{i\bk a\kQ}e^{+i\epsilon t} + \mathbf{Y}^{}_{i\bk a\kQ} e^{-i\epsilon t} \right ]
%\psi^*_{i\bk}(\mathbf{x})\psi_{a\kQ}(\mathbf{x}) .
%\label{eqn24}
%\end{equation}

Off-diagonal $\bB$ sub-blocks of the BSE matrix, which couple the resonant, $\bA$, and anti-resonant, $\mathbf{A}^*$, sub-blocks
of the matrix, typically have a limited effect on optical absorption properties predicted by the BSE \cite{Sander15}.
When this is the case, $\bY$ components of the eigenvectors in Eq. \ref{eqn21} are small compared to $\bX$ components.
The Tamm-Dancoff approximation (TDA) to the BSE omits the $\mathbf{B}$ sub-blocks in Eq. \ref{eqn21} and requires instead solution of 
the ordinary eigenvalue problems, 

\begin{eqnarray}
\label{eqn22}
\mathbf{A} \mathbf{X} &=& \mathbf{X} \mathbf{\Omega} \\
\nonumber
\mathbf{A}^* \mathbf{X}^* &=& -\mathbf{X}^* \mathbf{\Omega}.
\end{eqnarray}

The TDA is used from here on in this work. 

\subsection{Bare and dressed polarizabilities}

Resonant and anti-resonant parts of the non-interacting polarizability, $\Pi^0(\bQ,\epsilon)$, are usually obtained as products of
HF electron and hole Green's functions. They can also be obtained as the inverse of the BSE matrix in Eq. \ref{eqn21}.
For the case where $\bA$ and $\bB$ blocks are omitted from Eq. \ref{eqn21} so that only virtual and occupied energy eigenvalue
differences remain on the diagonal of the BSE matrix, $\bX$ is the unit matrix and $\bY$ is zero. 
Resonant and anti-resonant parts of the non-interacting HF polarizability, $\Pi^0(\bQ,\epsilon)$, are then given by,

\begin{eqnarray}
i\Pi^0_{Res.}(\bQ, \epsilon) = \sum_{i,a,\mathbf{k}}%^{occ,vir}
\frac{\psi^{}_{i\mathbf{k}}(\mathbf{x}_1)\psi^*_{a\kQ}(\mathbf{x}_1)
\psi^*_{i\mathbf{k}}(\mathbf{x}_2)\psi^{}_{a\kQ}(\mathbf{x}_2)}
{\epsilon - (\epsilon_{a\kQ} - \epsilon_{i\mathbf{k}}) + i\delta},
\label{eqn23}
\end{eqnarray}

\noindent
and, 

\begin{eqnarray}
i\Pi^0_{Anti-Res.}(\bQ, \epsilon) = - \sum_{i,a,\mathbf{k}}
\frac{\psi^{}_{a\mathbf{k}}(\mathbf{x}_1)\psi^*_{i\kQ}(\mathbf{x}_1)
\psi^*_{a\mathbf{k}}(\mathbf{x}_2) \psi^{}_{i\kQ}(\mathbf{x}_2)}
{\epsilon + (\epsilon_{a\mathbf{k}} - \epsilon_{i\kQ}) - i\delta}.
\label{eqn24}
\end{eqnarray}

\noindent
Similarly, the RPA-TDA polarizability, $\Pi(\mathbf{Q},\epsilon)$, can be expressed in terms of eigenvectors, $\bX$ and RPA-TDA eigenvalues, 
$\Omega^S(\mathbf{Q})$, which are obtained by solving the BSE equation using RPA-TDA $\bA$ matrix elements
in Table \ref{tab1}. Resonant and anti-resonant parts of the polarizability are obtained as inverses of the two subsystems
in Eq. \ref{eqn22}. The resonant part is obtained from the first of these equations and is given by,

\begin{widetext}

\begin{eqnarray}
i\Pi_{Res.}(\bQ, \epsilon) = \sum_{i,a,\bk,S}
\frac{\psi^{}_{i\mathbf{k}}(\mathbf{x}_1)\psi^*_{a\kQ}(\mathbf{x}_1)
\mathbf{X}^S_{i\bk a\kQ} \mathbf{X}^{S\dagger}_{i\bk a\kQ}
\psi^*_{i\mathbf{k}}(\mathbf{x}_2)\psi^{}_{a\kQ}(\mathbf{x}_2)}
%\frac{
%\psi_{a\kQ}(\mathbf{x}_1)\psi^*_{i\mathbf{k}}(\mathbf{x}_1)
%\mathbf{X}^\dagger_{a\kQ i\mathbf{k}} \mathbf{X}_{a\kQ i\mathbf{k}}
%                               \psi_{i\mathbf{k}}(\mathbf{x}_2)\psi^*_{a\kQ}(\mathbf{x}_2)}
{\epsilon - \Omega^S(\mathbf{Q}) + i\delta}
\label{eqn25}
\end{eqnarray}

\end{widetext}

\noindent
The RPA-TDA $\bA$ block matrix elements for the resonant polarizability are shown in Fig. \ref{fig2}(a).
The anti-resonant part can be obtained by setting up an additional RPA-TDA matrix corresponding to $\bA^*$ (Fig. \ref{fig2}(c)) 
and diagonalizing again. However, it is possible to obtain the non-resonant part using eigenvalues and eigenvectors
from the same $\bA$ matrix as the resonant part. Taking the complex conjugates of each wave function in the matrix element,

\begin{eqnarray}
(\psi^*_{a\kQ} \psi^{}_{i\bk} |V|\psi^*_{j\kq} \psi^{}_{b\kqQ})
\label{eqn26}
\end{eqnarray}

and using time-reversal symmetry, $\psi_\bk^* = \psi_{-\bk}$, yields,

\begin{eqnarray}
(\psi^*_{a-(\kQ)} \psi^{}_{i-\bk} |V|\psi^*_{j-(\kq)} \psi^{}_{b-(\kqQ)})
\label{eqn27}
\end{eqnarray}

\begin{figure}[htp]
\includegraphics[width=8.5cm,angle=0]{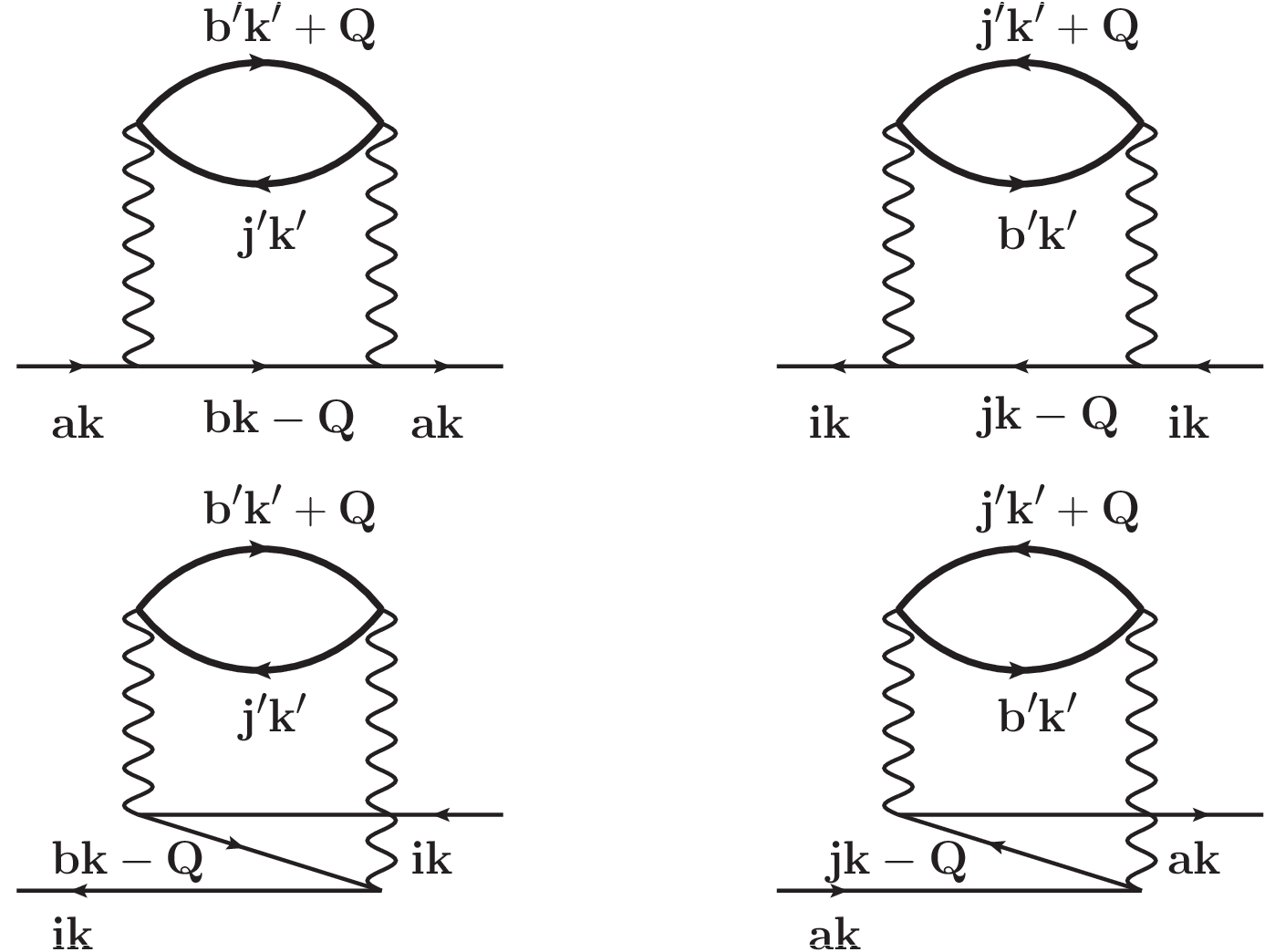}
\caption{Diagrams for matrix elements of the second order self-energy 
$\left <\psi_{n\mathbf{k}} | \Sigma^{(2)}(\bQ,\epsilon) | \psi_{n\mathbf{k}} \right >$.}
\label{fig4}
\end{figure}

\noindent
Adding a reciprocal lattice vector, $\bG$, to bring each wave vector into the first BZ yields,

\begin{eqnarray}
(\psi^*_{a\kppp} \psi^{}_{i\kppQ} |V|\psi^*_{j\kppqQ} \psi^{}_{b\kppq})
\label{eqn28}
\end{eqnarray}

\noindent
where,

\begin{eqnarray}
\label{eqn29}
\kppp &=& -(\kQ) + \bG\\
\nonumber
\bqpp &=& -\bq
\end{eqnarray}

\noindent
i.e. the RPA-TDA matrix $\bA^*$ has eigenvectors $\bX_{i^*\kQ a\bk}$ which are complex conjugates of eigenvectors of the RPA-TDA
matrix $\bA$ with the same eigenvalues. Wave vectors $\bk$ in $\mathbf{X}_{i\bk a\kQ}$ are reordered in $\mathbf{X}_{i^*\kQ a\bk}$.
Later, where the polarizability is used in the \gw self-energy, these eigenvectors are contracted with wave function products summed over
all wave vectors and thus this reordering is immaterial. The original RPA-TDA anti-resonant polarizability,

\begin{widetext}

\begin{eqnarray}
i\Pi_{Anti-Res.}(\bQ, \epsilon) = -\sum_{i,a,\bk,S}
\frac{\psi^{}_{a\mathbf{k}}(\mathbf{x}_1)\psi^*_{i\kQ}(\mathbf{x}_1)
\mathbf{X}^{S\dagger}_{i\kQ a\bk} \mathbf{X}^S_{i\kQ a\bk}
\psi^*_{a\mathbf{k}}(\mathbf{x}_2) \psi^{}_{i\kQ}(\mathbf{x}_2)}
%\psi^{}_{i\mathbf{k}}(\mathbf{x}_2)\psi^*_{a\kQ}(\mathbf{x}_2)}
%\frac{\psi_{i\mathbf{k}}(\mathbf{x}_1)\psi^*_{a\kQ}(\mathbf{x}_1)
%\mathbf{X}^\dagger_{a\kQ i\mathbf{k}} \mathbf{X}_{a\kQ i\mathbf{k}}
%\psi_{a\kQ}(\mathbf{x}_2)\psi^*_{i\mathbf{k}}(\mathbf{x}_2)}
{\epsilon + \Omega^S(\mathbf{Q}) - i\delta}
\label{eqn30}
\end{eqnarray}

\noindent 
can therefore be obtained as,

%\noindent
%Assuming time-reversal symmetry and making the substitution, $\bk' = -\bk -\bQ$, amplitudes,
%$\mathbf{X}^S_{i\kQ a\bk} \psi^{}_{a\mathbf{k}}(\mathbf{x})\psi^*_{i\kQ}(\mathbf{x})$,
%can be replaced by,
%$\mathbf{X}^{S\dagger}_{i\bk' a\mathbf{k'+Q}} \psi^{}_{i\bk'}(\mathbf{x})\psi^*_{a\mathbf{k'+Q}}(\mathbf{x})$,
%and the anti-resonant part of the polarizability becomes,

\begin{eqnarray}
i\Pi_{Anti-Res.}(\bQ, \epsilon) = -\sum_{i,a,\bk,S}
\frac{\psi^{}_{i\mathbf{k}}(\mathbf{x}_1)\psi^*_{a\kQ}(\mathbf{x}_1)
\mathbf{X}^S_{i\bk a\kQ} \mathbf{X}^{S\dagger}_{i\bk a\kQ}
\psi^*_{i\mathbf{k}}(\mathbf{x}_2)\psi^{}_{a\kQ}(\mathbf{x}_2)}
%\frac{\psi^*_{a\mathbf{k}}(\mathbf{x}_1)\psi^{}_{i\kQ}(\mathbf{x}_1)
%\mathbf{X}^\dagger_{i\kQ a\bk} \mathbf{X}^{}_{i\kQ a\bk}
%\psi^{}_{a\mathbf{k}}(\mathbf{x}_2) \psi^*_{i\kQ}(\mathbf{x}_2)}
%\frac{\psi^*_{a\mathbf{k}}(\mathbf{x}_1)\psi^{}_{i\kQ}(\mathbf{x}_1)
%\mathbf{X}^\dagger_{i\kQ a\bk} \mathbf{X}^{}_{i\kQ a\bk}
%\psi^{}_{a\mathbf{k}}(\mathbf{x}_2) \psi^*_{i\kQ}(\mathbf{x}_2)}
{\epsilon + \Omega^S(\mathbf{Q}) - i\delta}
\label{eqn31}
\end{eqnarray}

%\end{widetext}

\subsection{Self-energy}

The \gowo self-energy is the convolution of the non-interacting Green's function, $G_0(\br,\br',\epsilon)$, with a screened interaction
(Eq. \ref{eqn4}),

%\begin{widetext}

\begin{equation}
\Sigma^{GW}(\br,\br',\epsilon) = \int_{-\infty}^{+\infty} \frac{d\epsilon'}{2 \pi} %e^{-i\delta \epsilon} 
iG_0(\br, \br', \epsilon - \epsilon') W_0(\br, \br', \epsilon').
\label{eqn32}
\end{equation}

\noindent
The bare coulomb interaction, $\vQ$, in Eq. \ref{eqn4} contributes the static, HF exchange part of the self-energy, which is included 
in the HF SCF calculation in this work. The second order self-energy, $\Sigma^{(2)}(\br,\br',\epsilon)$, is obtained using the 
non-interacting polarizability (Eq. \ref{eqn24} and \ref{eqn25}) in Eq. \ref{eqn4}. Diagonal matrix elements of the second order self-energy, 
$\Sigma^{(2)}(\bQ, \epsilon)$ (Fig. \ref{fig4}), are given by, 

\begin{eqnarray}
\left <\psi^{}_{n\kp} | \Sigma^{(2)}(\epsilon) | \psi^{}_{n\kp} \right > = \sum_{j,b,j',b',\kpp,\bQ}
\frac{
( \psi^{}_{n\kp} \psi_{b\kmQ}^*     | \psi^{}_{j'\kpp} \psi_{b'\kpQp}^* )
( \psi_{j'\kpp}^* \psi^{}_{b'\kpQp} | \psi^{}_{b\kmQ} \psi_{n\kp}^* )
} {\epsilon_{} - \epsilon_{b\kmQ} - (\epsilon_{b'\kpQp} - \epsilon_{j'\kpp})}
+
\frac{
( \psi^{}_{n\kp} \psi_{j\kmQ}^*     | \psi^{}_{b'\kpp} \psi_{j'\kpQp}^* )
( \psi_{b'\kpp}^* \psi^{}_{j'\kpQp} | \psi^{}_{j'\kmQ} \psi_{n\kp}^* )
} {\epsilon_{} - \epsilon_{j\kmQ} - (\epsilon_{j'\kpQp} - \epsilon_{b'\kpp})}.
\label{eqn33}
\end{eqnarray}

\noindent
The \go self-energy is obtained using the interacting polarizability (Eq. \ref{eqn30} and \ref{eqn31}) in Eq. \ref{eqn4}. 
Diagonal matrix elements of $\Sigma^{GW}(\br,\br',\epsilon)$ are given by,

\begin{eqnarray}
\label{eqn34}
\left <\psi^{}_{n\kp} | \Sigma^{GW}(\epsilon) | \psi^{}_{n\kp} \right > = \sum_{j,b,j',b',\kpp,\bQ,S}
\frac{
( \psi^{}_{n\kp} \psi_{b\kmQ}^*     | \psi^{}_{j'\kpp} \psi_{b'\kpQp}^* )
\mathbf{X}^S_{j'\kpp b'\kpQp} \mathbf{X}^{S\dagger}_{j''\kppp b''\kppQp}
( \psi_{j'\kpp}^* \psi_{b'\kpQp} | \psi_{b\kmQ} \psi_{n\kp}^* )
} {\epsilon - \epsilon_{b\kmQ} - \Omega^S(\bQ)}
+\\
\nonumber
\frac{
( \psi_{n\kp} \psi_{j\kmQ}^*     | \psi_{b'\kpp} \psi_{j'\kpQp}^* )
\mathbf{X}^S_{j'\kpp b'\kpqp} \mathbf{X}^{S\dagger}_{j''\kppp b''\kppp}
( \psi_{b'\kpp}^* \psi_{j'\kpQp} | \psi_{j'\kmQ} \psi_{n\kp}^* )
} {\epsilon - \epsilon_{j\kmQ} + \Omega^S(\bQ)}
\end{eqnarray}

\end{widetext}

\noindent
Various self-energy approximations are possible within this approach by solving particular approximations to Eq. \ref{eqn21} 
and using the resulting eigenvectors and excitation energies to construct the self-energy.
The computational expense of the RPA-TDA calculations used to construct the self-energy depends on the number of occupied and virtual 
states used. Most of the self-energy for valence excitations is obtained from transitions between states close to the Fermi level. 
Here we use Eq. \ref{eqn34} to recover most of the self-energy and Eq. \ref{eqn33} for the remainder. The latter does not require a 
RPA-TDA calculation and is therefore less expensive to evaluate. For example, for diamond, four occupied states and 10 virtual states 
were included in the \gowo self-energy and transitions between these occupied states and an additional 20 states were included in the 
second order self-energy. The RPA-TDA matrix for an 8x8x8 Monkhorst-Pack (MP) mesh with these occupied and virtual states is of dimension 
20480 and for a 10x10x10 mesh it is of dimension 40000.

\subsection{Density fitting}

\begin{figure}[htp]
\includegraphics[width=2cm,angle=0]{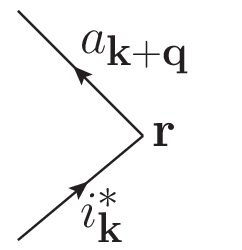}
\caption{Charge density at point, $\mathbf{r}$, arising from wavefunction product $\psi^*_{i\mathbf{k}}(\mathbf{r}) 
\psi_{a\kq}(\mathbf{r})$.}
\label{fig5}
\end{figure}

Density fitting is used to project wavefunction products,

\begin{eqnarray*}
\psi^*_{a\mathbf{k}}(\mathbf{r}) \psi_{b\kq}(\mathbf{r}), \\
\nonumber
\psi^*_{a\mathbf{k}}(\mathbf{r}) \psi_{i\kq}(\mathbf{r}), \\
\nonumber
\psi^*_{i\mathbf{k}}(\mathbf{r}) \psi_{a\kq}(\mathbf{r}), \\
\nonumber
\psi^*_{i\mathbf{k}}(\mathbf{r}) \psi_{j\kq}(\mathbf{r}),
\end{eqnarray*}

\noindent
i.e. virtual-virtual, virtual-occupied, occupied-virtual and occupied-occupied products at wave vectors $\bk$ and $\bkq$, onto an 
auxiliary Gaussian orbital basis. Bloch functions with lattice translation symmetry are constructed from crystal orbitals (CO),

\begin{equation}
\psi_{i\bk}(\br) = c_{i\bk m} \phi_m(\br - \bmp - \bA)e^{i\bk.\bA},
\label{eqn35}
\end{equation}

\noindent
which are linear combinations of basis functions, $\phi_m(\br - \bmp - \bA)$, at site, $\bmp$, in the unit cell with lattice 
translation vector, $\bA$, and expansion coefficient, $c_{i\bk m}$, for the $i^{th}$ occupied state at wavevector, $\bk$.
Wavefunction product densities (Fig. \ref{fig5}) are expressed in auxiliary basis functions, 
$\chi_\beta^{+\mathbf{q}}(\mathbf{r})$, as \cite{Patterson20a},

\begin{eqnarray}
\psi^*_{i\mathbf{k}}(\mathbf{r}) \psi^{}_{a\kq}(\mathbf{r}) = 
\left ( \psi^*_{i\mathbf{k}} \psi^{}_{a\kq} | \chi^{-\mathbf{q}}_\alpha \right ) \left[ V_{\alpha\beta}^{\mathbf{-q}}\right]^{-1} 
\chi_\beta^{+\mathbf{q}}(\mathbf{r}).
\label{eqn36}
\end{eqnarray}

\noindent
The auxiliary basis CO, $\chi_\beta^{+\mathbf{q}}(\br)$, is a lattice sum of auxiliary basis functions on site, $\bn$, in 
unit cell, $\bA$,

\begin{equation}
\chi_\alpha^{+\mathbf{q}}(\br) = \chi_\alpha(\br - \bn - \bA)e^{i\bq.\bA}.
\label{eqn37}
\end{equation}

\noindent
$\left ( \psi^*_{i\mathbf{k}} \psi^{}_{a\kq} | \chi^{-\mathbf{q}}_\alpha \right ) $ is the three-center integral,

\begin{equation}
V_{mn\beta}^{\bk,\bq} = \sum_{\bB,\bC} \drdrp \frac{ \phi_{m}^{*}(\br) \phi_{n}(\br - \bC) \chi_{\beta}^{*}(\brp) } {\ewaldb} e^{i(\bkq).\bC} e^{-i\bq.\bB},
\label{eqn38}
\end{equation}

\noindent
contracted with expansion coefficients of the wavefunctions, $\psi^*_{i\mathbf{k}}(\br)$ and $\psi^{}_{a\kq}(\br)$, and 
$\left[ V_{\alpha\beta}^{\mathbf{-q}}\right]^{-1}$ is the inverse of the matrix,

\begin{eqnarray}
V_{\alpha\beta}^{-\bq} = \sum_{\bA} \drdrp \frac{\chi_{\alpha}(\br) \chi_{\beta}^{*}(\brp) } {\ewalda}  e^{-i\bq.\bA}.
\label{eqn39}
\end{eqnarray}

\noindent
Both are integrals over the Ewald potential,

\begin{widetext}

\begin{equation}
\sum_{\bA} \frac{e^{i\bq.\bA}}{\ewalda} = \sum_{\bG} \frac{4\pi}{\Omega} \frac{e^{-\frac{|\bq + \bG|^2}{4\gamma}}}{|\bq + \bG|^2} 
e^{i(\bq + \bG).(\br - \brp)} + \sum_{\bA} \frac{\text{erfc}(\gamma^{1/2}|\br - \brp - \bA|)}{|\br - \brp - \bA|} e^{i\bq.\bA}.
\label{eqn40}
\end{equation}

\noindent
Combining two factors of density products in Eq. \ref{eqn36} and the Ewald potential, results in the following expression 
\cite{Patterson20a} for the leftmost, unscreened two-electron integral in Fig. \ref{fig2}(a),

\begin{eqnarray}
( \psi^*_{a\kQ}(\br) \psi^{}_{i\bk}(\br) | \psi^*_{j\bkq}(\brp) \psi^{}_{b\kqQ}(\brp) ) = 
( \psi^*_{i\bk} \psi^{}_{a\kQ} | \chi^{-\bQ}_\alpha )^* \left[ V_{\alpha\beta}^{\mathbf{-Q}} \right]^{-1}
( \psi^*_{j\bkq} \psi^{}_{b\kqQ} | \chi^{-\bQ}_\beta ).
\label{eqn41}
\end{eqnarray}

\noindent
Further details of the density fitting procedure are given in Ref. [\onlinecite{Patterson20a}]. 

The \gowo self-energy in Eq. \ref{eqn34} contains two two-electron integrals contracted with $\mathbf{X}$ amplitudes from $\bQ$-dependent 
RPA-TDA calculations and a denominator consisting of single-particle energies, $\epsilon_{j\kmQ}$ or $\epsilon_{b\kmQ}$, 
and RPA-TDA energies, $\Omega^S(\bQ)$. The numerator of the first term on the $rhs$ of Eq. \ref{eqn34} translated into the two- and 
three-center integrals of Eq. \ref{eqn30} and \ref{eqn31} and RPA-TDA eigenvectors, $\mathbf{X}^S_{j'\kpp b'\kpQp}$, becomes,

\begin{eqnarray}
( \psi^{}_{n\kp} \psi_{b\kmQ}^*  | \chi_\alpha^{-\bQ} )
\left[ V_{\alpha\beta}^{\bQ-1} \right]^{*-1}
( \chi^{+\bQ}_\beta              | \psi^{}_{j'\kpp} \psi_{b'\kpQp}^* )
\mathbf{X}^S_{j'\kpp b'\kpQp} \mathbf{X}^{S\dagger}_{j''\kppp b''\kppQp}
( \psi_{j'\kpp}^* \psi_{b'\kpQp} | \chi^{-\bQ}_\delta )
\left[ V_{\delta\gamma}^{\bQ-1} \right]^{*-1}
( \chi^{+\bQ}_\gamma             | \psi_{b\kmQ} \psi_{n\kp}^* )
\label{eqn42}
\end{eqnarray}

\end{widetext}

\noindent
While this expression might seem cumbersome, it is relatively straightforward to assemble. First, the RPA
eigenvectors are contracted with three-center integrals to form products,

\begin{eqnarray}
X_{S,\beta} = ( \chi^{+\bQ}_\beta | \psi^{}_{j'\kpp} \psi_{b'\kpQp}^* ) \mathbf{X}^S_{j'\kpp b'\kpQp}
\label{eqn43}
\end{eqnarray}

\noindent
Second, these products are contracted with,

\begin{eqnarray}
( \psi^{}_{n\kp} \psi_{b\kmQ}^*  | \chi_\alpha^{-\bQ} ) \left[ V_{\alpha\beta}^{\bQ-1} \right]^*
\label{eqn44}
\end{eqnarray}

\noindent
to yield the factor on the $lhs$ of Eq. \ref{eqn42}. For diagonal elements of the self-energy operator, the factor on the $rhs$ is
simply the complex conjugate of the $lhs$ factor. In practice, parallel calculation of three-center integrals is distributed over 
auxiliary basis set centers, $\alpha$, and RPA-TDA eigenvectors, $\mathbf{X}^S_{j'\kpp b'\kpQp}$, are distributed
over cores in a block-cyclic format following diagonalization using the ELPA package \cite{Auckenthaler11,Marek14}. 
The first step is completed by circulating RPA-TDA eigenvectors and multiplying them by the
three-center integrals on each core. This is followed by contraction over auxiliary basis function index, $\alpha$, to yield
the left half of the numerator in Eq. \ref{eqn42}. The self-energy matrix is assembled using an \textit{MPI-reduce} call to sum 
self-energy contributions over auxiliary basis function sites.

\subsection{Screened Interaction}

The bare Coulomb interaction in this work is represented by the Coulomb matrix in Eq. \ref{eqn39}, which is in the auxiliary basis.
The auxiliary basis is larger than the wavefunction basis and is of order 100 functions per atom. A matrix representing
a unit cell with 100 atoms is therefore of dimension around 10,000. Inverses of the Coulomb matrix 
arise in representation of wavefunction product densities in Eq. \ref{eqn36}; a matrix of this dimension is relatively 
inexpensive to invert. 

Screened interactions in BSE matrix element calculations are approximated using a static, inverse dielectric matrix \cite{Hybertsen86}. 
In this work, a static approximation to the screened interaction is obtained by omitting the energy dependence of the dressed 
polarizability in Eq. \ref{eqn30} and \ref{eqn31} and combining resonant and anti-resonant terms into a single term. 
Products of densities in the polarizability are expressed as combinations of three-center integrals and inverted Coulomb matrices
(Eq. \ref{eqn36}). The matrix representation of the static, dressed polarizability is,

\begin{widetext}

\begin{eqnarray}
i \left[ \Pi^{-\bq}_{\alpha\beta} \right]^{-1} =
4 \frac{
\left[ V_{\alpha\beta}^{\bq-1} \right]^*
( \chi^{+\bq}_\beta              | \psi^{}_{j'\kpp} \psi_{b'\kpqp}^* )
\mathbf{X}^S_{j'\kpp b'\kpqp} \mathbf{X}^{S\dagger}_{j''\kppp b''\kppqp}
( \psi_{j'\kpp}^* \psi_{b'\kpqp} | \chi^{-\bq}_\delta )
\left[ V_{\delta\gamma}^{\bq-1} \right]^*
} {\Omega^S(\bq)}.
\label{eqn45}
\end{eqnarray}

\noindent
where the factor of 4 arises from combining the resonant and anti-resonant terms in the polarizability and summing over spin. 
The screening part of $W_0(\bQ,\epsilon)$ is obtained by multiplying this polarizability on the left and right by the density products
$\psi^*_{a\kQ}(\br) \psi^{}_{b\kqQ}(\br)$ and $\psi^*_{j\bkq}(\brp) \psi^{}_{i\bk}(\brp)$ and integrating over the CO auxiliary
basis functions which transform Eq. \ref{eqn45} into its coordinate representation to give,

\begin{eqnarray*}
\nonumber
( \psi^*_{b\kqQ} \psi^{}_{a\kQ} | \chi^{-\bq}_\alpha )^* \left[ i \Pi_{\alpha\beta}^{\mathbf{-q}} \right]^{-1}
( \chi^{-\bq}_\beta | \psi^*_{i\bk} \psi^{}_{j\bkq} ).
\end{eqnarray*}

The total screened interaction (Eq. \ref{eqn4}) is a sum of the bare Coulomb interaction plus this screening part with
matrix representation,

\begin{eqnarray}
\left[ W_{\alpha\beta}^{-\bq}\right]^{-1} = \left[ V_{\alpha\beta}^{-\bq} \right]^{-1} + \left[i\Pi^{-\bq}_{\alpha\beta} \right]^{-1}
\label{eqn46}
\end{eqnarray}

\noindent
Once this has been obtained, a range of screened interaction matrix elements (Table \ref{tab1}) can be obtained by multiplying
the appropriate three-center integrals on the right and left. For example, diagram (a) in Fig. \ref{fig2} is given by,

\begin{eqnarray}
\label{eqn47}
( \psi^*_{a\kQ}(\br) \psi^{}_{b\kqQ}(\br) | W | \psi^*_{j\bkq}(\brp) \psi^{}_{i\bk}(\brp) ) = 
( \psi^*_{b\kqQ} \psi^{}_{a\kQ} | \chi^{-\bq}_\alpha )^* \left[ W_{\alpha\beta}^{\mathbf{-q}} \right]^{-1}
( \chi^{-\bq}_\beta | \psi^*_{i\bk} \psi^{}_{j\bkq} ).
\end{eqnarray}

\end{widetext}

%\begin{eqnarray}
%( \psi^*_{a\kQ}(\br) \psi^{}_{b\kqQ}(\br) | W |  \psi^*_{j\bkq}(\brp) \psi^{}_{i\bk}(\brp) ) = 
%( \psi^*_{b\kqQ} \psi^{}_{a\kQ} | \chi^{-\bq}_\alpha )^* \left[ \W_{\alpha\beta}^{\mathbf{-q}} \right]^{-1}
%( \chi^{-\bq}_\beta | \psi^*_{i\bk} \psi^{}_{j\bkq} ).
%\label{eqn39}
%\end{eqnarray}
%The density products are expressed in terms of three center integrals and Coulomb matrix inverses. 
%The product of Coulomb matrix inverses and the Coulomb matrix at wave vector $\bq$ yield the identity so that 
%Comparing Eq. \ref{eqn39} to  Eq. \ref{eqn33}, we see that a total screened interaction can be expressed as, 

%Screened and unscreened Coulomb interactions as a function of wavevectors $\mathbf{q}$ and $\mathbf{Q}$ are shown in Fig. \ref{fig2}
%and the meanings of these wave vectors were shown in Fig. \ref{fig1}. Diagrams (b) and (d) in Fig. \ref{fig2} have screened interactions 
%with momentum transfer, $\bq$, whereas diagrams (a) and (c) have unscreened Coulomb interactions with momentum transfer, $\bQ$. 
%Calculation of matrix elements of the unscreened Coulomb interaction in Fig. \ref{fig2} requires integration of the product
%densities with the unscreened Coulomb potential. The result is a product of two three-center integrals and the inverted
%Coulomb matrix \ref{eqn33}.

\subsection{Small $\mathbf{Q}$ limits\label{smallq}}

The leading terms in the contribution to the self-energy and statically screened electron-hole interaction in the limit as 
$\bQ \to \mathbf{0}$ in Eq. \ref{eqn34} and and Eq. \ref{eqn45} arise from
forward scattering of the electron or hole in the two diagrams at the top of Fig. \ref{fig4} and diagrams (a) and (c) in Fig. \ref{fig3}. 
The numerator of the first term in the \gowo self-energy in Eq. \ref{eqn34} was translated into its density-fitted expression in 
Eq. \ref{eqn42}. This is divergent as $\mathbf{Q} \to \mathbf{0}$ for the $\mathbf{G} = \mathbf{0}$ term in the Ewald potential 
in Eq. \ref{eqn40}. In order to isolate the $Q^2$ divergence, the first term on the $rhs$ of Eq. \ref{eqn34} is rewritten in its 
Fourier representation as,

\begin{eqnarray}
%\sum_{j,b,j',b',\kpp,S}
2 \frac{4 \pi}{Q^2} |(\psi^{}_{n\kp} | \mbQr | \psi_{b\kmQ}^*)|^2
\frac{4 \pi}{Q^2} 
\frac{| \mathbf{X}^S_{j'\kpp b'\kpQp} ( \psi^{}_{j'\kpp}| \bQrp | \psi_{b'\kpQp}^* ) |^2} {( \epsilon - \epsilon_{b\kmQ} - \Omega^S(\bQ))},
\label{eqn48}
\end{eqnarray}

\noindent
where repeated indices are summed over and a factor of 2 is included for spin summation in the polarizability.
The leading contributions to the first factor come from $\psi^{}_{n\kp}=\psi^{}_{b\kp}$; the matrix element of $\bQr$ approaches 
unity as $\mathbf{Q} \to \mathbf{0}$. Matrix elements with other wavefunctions, $\psi^{}_{n\kp} \neq \psi^{}_{b\kp}$, at that order vanish
by orthogonality. Leading contributions to the second factor are found at first order in $\bQrp$: 
$i\bQ . ( \psi^{}_{j'\kpp}| \brp | \psi_{b'\kpQp}^* )$. This is evaluated using \cite{Lin77},

\begin{eqnarray}
i\left< 0 | \mathbf{p} | S \right> = \left< 0 | \left[ \mathbf{r}, H^{RPA} \right] | S \right> = \Omega^S(\bQ) 
 \left< 0 | \mathbf{r} | S \right>,
\label{eqn49}
\end{eqnarray}

\noindent
where 0 and \textit{S} are the ground and S$^{th}$ RPA excited states, respectively, and $H^{RPA}$ is the RPA Hamiltonian. 
It is well known that the non-local HF exchange operator in the single-particle Fock operator does not commute with the position 
operator \cite{Starace71} leading to an extra term in the commutator above when the Hamiltonian is the Fock operator. 
However, the RPA Hamiltonian commutes with the position operator \cite{Lin77}, hence the commutator of the exchange and position
operators does not appear here.

In our notation the contraction of the RPA eigenvector for state \textit{S}, $\mathbf{X}^S_{j'\kpp b'\kpQp}$, with the position matrix 
element in Eq. \ref{eqn49} becomes,

\begin{eqnarray}
%\sum_{j'\kpp b'\kpQp} ( \psi^{}_{j'\kpp}| \brp | \psi_{b'\kpQp}^* )\mathbf{X}^S_{j'\kpp b'\kpQp} = \sum_{j'\kpp b'\kpQp} \frac{i( \psi^{}_{j'\kpp}| \mathbf{p} | \psi_{b'\kpQp}^* ) \mathbf{X}^S_{j'\kpp b'\kpQp} }{\Omega^S(\bQ)}.
( \psi^{}_{j'\kpp}| \brp | \psi_{b'\kpQp}^* )\mathbf{X}^S_{j'\kpp b'\kpQp} = 
i\frac{( \psi^{}_{j'\kpp}| \mathbf{p} | \psi_{b'\kpQp}^* ) \mathbf{X}^S_{j'\kpp b'\kpQp} }{\Omega^S(\bQ)},
\label{eqn50}
\end{eqnarray}

\noindent
The contribution to the self-energy of virtual state $\psi_{n\bk} = \psi_{b\kmQ}$ with energy 
$\epsilon = \epsilon_{b\kmQ}$ in the limit of small $\mathbf{Q}$ is,

\begin{eqnarray}
%\sum_{j,b,j',b',\kpp,S}
- \frac{1}{N_{\bQ}N_{\kpp}}\frac{16 \pi^2}{Q^2} 
\frac{| \hat{\bQ} . ( \psi^{}_{j'\kpp}| \mathbf{p} | \psi_{b'\kpQp}^* ) \mathbf{X}^S_{j'\kpp b'\kpQp} |^2 } 
{\Omega^{S}(\bQ)^3}.
\label{eqn51}
\end{eqnarray}

\noindent
where $\hat{\bQ}$ is a unit vector parallel to $\bQ$ and $N_{\bQ}$ and $N_{\kpp}$ are the number of $\bQ$ and $\kpp$ wave vectors 
in the Brillouin zone. There is a similar, positive term corresponding to the second term on the
right in Eq. \ref{eqn34} for valence states. The $Q^2$ divergence is treated using the method of Gygi and Baldereschi \cite{Gygi86} 
for handling this divergence in exchange energies of cubic solids; otherwise the extension of this method to
general lattices by Carrier, Rohra and G{\"o}rling \cite{Carrier07} is used. The form given in Eq. \ref{eqn51} is used for the
non-analytic, divergent part as $\bQ \to \mathbf{0}$ and the non-divergent, density-fitted form in Eq. \ref{eqn34} is used for finite $\bQ$.
The main uncertainty in this approach to the small $\bQ$ limit is the equivalence of momentum and position matrix elements implied 
by Eq. \ref{eqn49} in an incomplete basis. Similar divergences in the screened electron-hole attraction (Eq. \ref{eqn47}) are treated 
in the same way.

\section{Computational Methods\label{comp}}

Basis sets in this work were all electron GO basis sets adapted from the def2-TZVP basis sets of Weigend and Ahlrichs
\cite{Weigend05}. Gaussian orbital basis sets are commonly generated using the variational principle, i.e. minimization of
total energy of the system in its ground state. However, response properties require conduction bands to be well
represented as well as occupied states. The completeness of the basis sets used in the kinetic energy range to 100 eV
were tested by diagonalizing the kinetic energy operator and generating free-electron band structures for the materials
studied. These are shown in the supporting information. This was done for the def2-TZVP basis sets and for augmented basis sets
with modified exponents and additional high angular momentum basis functions. Diffuse basis functions
(orbital exponent $\alpha \leq 0.1$) cause linear dependence problems and are not essential to construct free-electron
band structures up to 100 eV. 

For MgO, the def2-TZVP Mg(O) basis sets have 4s3p2d(5s3p2d1f) functions (with diffuse functions excluded).
Deviations from the free-electron band structure for MgO are of order several eV for kinetic energies as low as 20 eV (Fig. S1).
However, augmenting this to 6s4p2d1f1g for both Mg and O results in a free-electron band structure which shows deviations from
the free electron band structure around 60 eV, but generally is much improved over the def2-TZVP basis for energies of 20 to 100 eV.

For diamond, the def2-TZVP C 11s5p2d1f(5s3p2d1f) basis set was replaced by the 11s6p2d2f(5s4p2d2f) basis given in SI Table S1. The 
def2-TZVP basis set core functions were retained and diffuse s and p functions were replaced by less diffuse functions. The free
electron band structure is reproduced well by this basis up to 120 eV (SI Fig. S1).

For Si, the 13s9p1d(6s5p1d) basis from Heyd and coworkers \cite{Heyd05} was replaced by the 13s9p3d1f(6s5p3d1f) basis given in SI Table S2. 
Changes to the original basis were to add d and f functions.

For anatase and rutile TiO$_2$ the def2-TZVP 17s11p7d1f(6s4p4d1f) basis was replaced by the 16s11p7d2f(5s4p3d2f) basis given in
SI Table S3. Diffuse s, p and d functions were replaced by less diffuse functions. The free electron band structure for the TiO$_2$ 
phases is reproduced well up to about 60 eV. The O basis for the TiO$_2$ phases differed from the O basis for MgO in the number 
and exponents of d and f functions.

Auxiliary basis sets used for density fitting were the def2-TZVP-RIFIT sets \cite{Weigend98}. Basis functions of h or higher angular
momentum were omitted. Modified wavefunction basis sets are listed in Supporting Information Tables S1 to S5.

\section{Results\label{results}}

The materials chosen for study here are those used in a previous study of the performance of TDHF in wide gap materials with a scaled
electron-hole attraction \cite{Patterson20a}, namely diamond C, MgO and anatase and rutile TiO$_2$. Here we also consider
Si as an example of a narrow, indirect gap semiconductor with a valence band of $sp^3$ character. HF theory
overestimates band widths of materials where the bottom and top of the band are of different orbital character. For example,
the bottom and top of the valence band in diamond and Si are of $s$ and $p$ character, respectively, and HF theory overestimates
the band width in diamond C and Si by about 25\%. Below we show that \goHF corrects this and predicts band widths
in agreement with experiment. In materials such as the oxides studied here the O 2$s$ and 2$p$ bands are distinct.
In this case the HF O 2$p$ bands are nearly indistinguishable from the \goHF bands once both band structures have their VBM
aligned. The O 2$s$ levels shift upwards relative to the valence $p$ bands on going from HF to \goHF. 
There is a large renormalisation of the HF band gap in all these materials. \goHF band gaps are larger than \goLDA gaps and,
typically, experimental gaps lie between the theoretical predictions. 

\begin{table}[ht!]
\caption{Occupied and virtual state ranges and energy cutoff in eV used in \goHF and BSE@HF calculations and MP
meshes and dimensions for RPA screening calculations.}
\begin{ruledtabular}
\begin{tabular}{lcccccc}
Material        & Occupied & Virtuals$^a$ & Virtuals$^b$ & MP-net   & RPA$^c$ \\
\hline                                                                                                           
C               &    4     & 10 (39)      & 24 (110)     & 10x10x10 & 40000   \\
Si              &    4     & 14 (34)      & 32 (58)      & 10x10x10 & 56000   \\
MgO             &    4     & 14 (56)      & 24 (107)     & 10x10x10 & 56000   \\
Anatase TiO$_2$ &   12     & 20 (30)      & 2  (33)      & 6x6x6    & 51840   \\
Rutile  TiO$_2$ &   12     & 20 (30)      & 2  (33)      & 6x6x6    & 51840   \\
\end{tabular}
\begin{tablenotes}
\item[a] $^a$ RPA screening calculation number of virtual states (cutoff in eV) \\
\item[b] $^b$ Second-order $\Sigma$ calculation number of virtual states (cutoff in eV) \\
\item[c] $^c$ RPA screening matrix dimension \\
\end{tablenotes}
\end{ruledtabular}
\label{tab2}
\end{table}

In a conventional PW \go calculation wave functions are expanded in PW basis sets. PW cutoff energies of 230 and 680 eV for
wave functions in diamond and Si were used in early \goPWLDA calculations \cite{Hybertsen85} and found to be converged. 
RPA dielectric matrices for diamond and Si with dimension 220 x 220 and 140 x 140 $\bG$ vectors were used in the same work.
Shell closings around 220 $\bG$ vectors in diamond and 140 vectors in Si correspond to free electron energies of 32 and 10 eV, respectively.
Thus RPA excitations which are likely to be important in screening are relatively low in energy.
Tests of virtual state cutoff in RPA screening in molecular tetracene (C. H. Patterson unpublished) showed that a virtual state 
cutoff energy of 60 eV results in a difference in low energy excited states of less than 10 meV in a cc-pVTZ basis, compared to 
inclusion of all virtual states, and that convergence is essentially reached by including virtual states up to 50 eV. 
GO basis sets used in this work in \goHF calculations for diamond reproduce the free electron dispersion relation accurately to 
about 120 eV, when electron-electron and electron-nuclear interactions are turned off (SI Fig. S3). 
GO in which wave functiopns are expanded have radial decay exponents which correspond to electron kinetic energies well above 120 eV. 
Similarly, the PW cutoff energies used for wave functions in diamond or Si are well above this energy. 
However, in practice in both PW and GO calculations cutoff energies for the screening response do not need to exceed 100 eV.

Table \ref{tab2} shows the number of occupied and virtual states included in the \goHF (Eq. \ref{eqn34}) and second-order
(Eq. \ref{eqn33}) self-energies. For diamond, a virtual state cutoff energy of 39 eV is reached with just 10 conduction bands
in the RPA calculation used in the self-energy and screened interaction (Eq. \ref{eqn22}).
A further 24 conduction bands used in the second-order self-energy results in a cutoff energy of 110 eV.
For Si, 14 and 32 virtual states in each self-energy result in lower cutoffs of 34 eV in the RPA calculation and 58 eV in the 
second order self-energy, but well above the cutoff energy of 10 eV implied by 140 $\bG$ vectors in the PW RPA matrix. For MgO, 14 and 
24 virtual states in the respective self-energies resulted in cutoffs of 56 and 107 eV.

For the two TiO$_2$ phases, the cutoff in the RPA calculation is 30 eV for an RPA matrix size of 51840 with 40 (rutile) and 30
(anatase) unique $\bQ$ points at which the RPA matrix must be diagonalized. To achieve a cutoff of 50 eV in the RPA screening calculations
would require 44 virtual states for rutile TiO$_2$ and 48 virtual states for anatase TiO$_2$ and RPA matrix dimensions of 
114048 and 124416, which would need to be diagonalized at 40 and 30 unique $\bQ$ points for a 6x6x6 MP mesh. 
This is beyond our computational resources.

\section{$GW$ Band Structure}

\subsection{Diamond}

The band structure of diamond has been investigated by photoemission \cite{Himpsel80,Jimenez97,Edmonds13}, optical absorption
\cite{Roberts67} and electron energy loss spectroscopy \cite{Korneychuk18}. The \gowo band structure of diamond has been calculated
starting from LDA \cite{Rohlfing93,Lofas11,Edmonds13,Gao15,Nabok16} in GO \cite{Rohlfing93}, LAPW \cite{Nabok16} and PW \cite{Lofas11,Gao15}
basis sets. Valence and conduction band energies for diamond at high symmetry points of the BZ in these calculations are compared
to our HF starting point calculations in a GO basis in Table \ref{tab3}. The valence band width of diamond has been reported
to be between 21 eV \cite{Himpsel80} and 24.2 eV \cite{McFeely74}, with most experimental estimates lying between 23 and 24 eV
\cite{Edmonds13}. 

\begin{figure}[htp]
\includegraphics[width=6cm,angle=0]{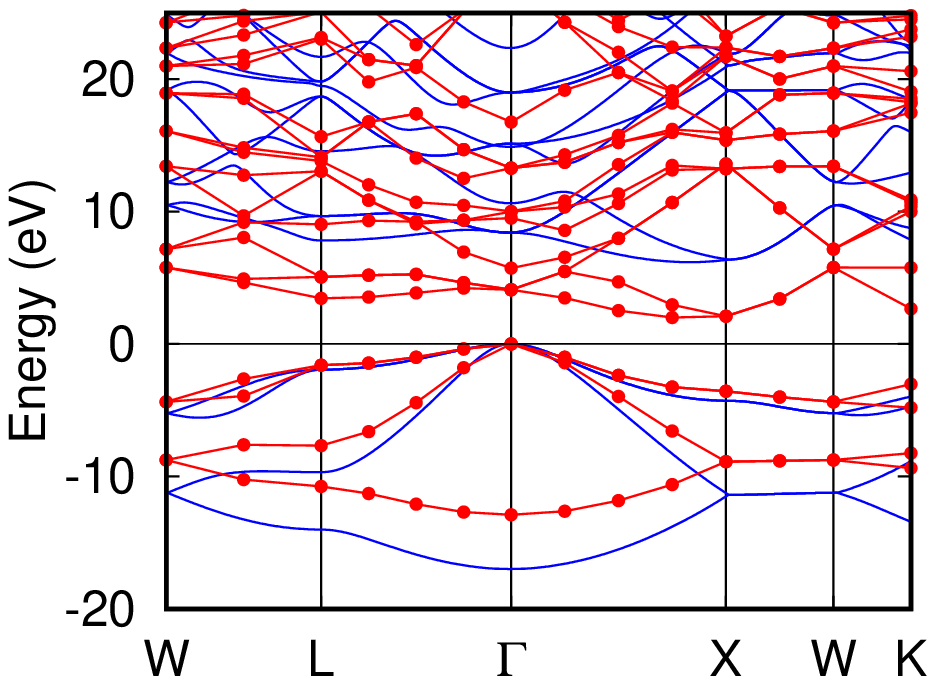}
\includegraphics[width=6cm,angle=0]{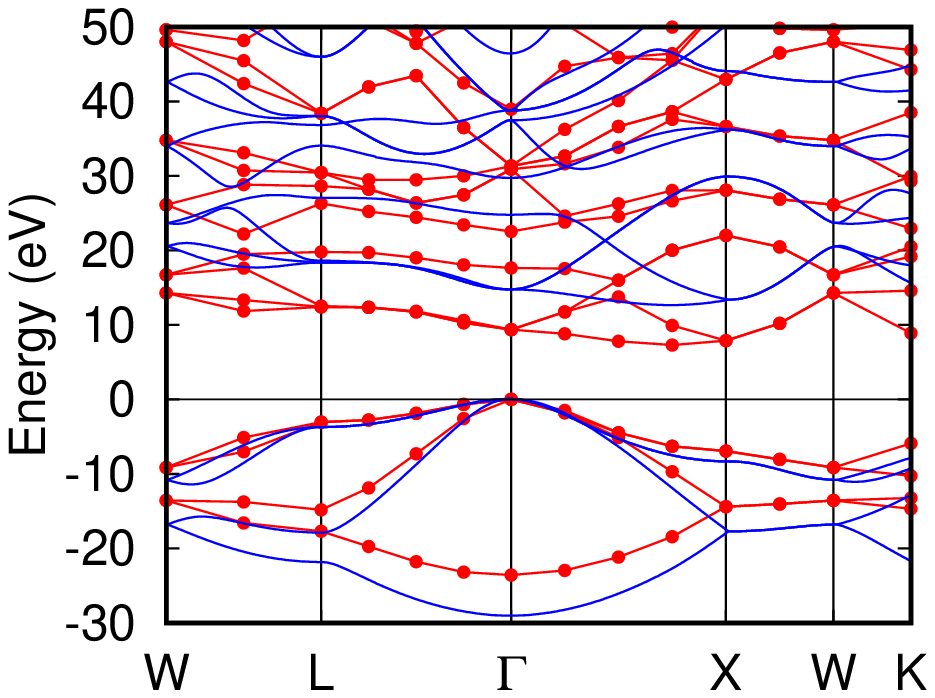}
\caption{Band structures of Si and C in the diamond structure from HF and \gowo calculations. HF band structures are shown as solid blue 
lines. \gowo self-energies were obtained in an 8x8x8 MP mesh at points shown.}
\label{fig6}
\end{figure}

\begin{table*}
\caption{Valence and conduction band energies for diamond at high symmetry points in the BZ from HF, \goHF and \goLDA and photoemission 
experiments. All values are relative to the VBM at the $\Gamma$ point. All \gowo energies, with the exception of this work, used a 
LDA starting point and GO, LAPW or PW basis sets. This work used a HF starting point and a GO basis set.}
\begin{ruledtabular}
\begin{tabular}{lccccccccc}
State           & HF$^a$ &\goHFp$^a$&\goLDA$^b$&\goLDA$^c$&\goLDA$^d$&\goLDA$^e$ &\goLDA$^f$& Expt$^f$            & Expt             \\
\hline                                                                                                                           
$\Gamma_{c}$    &  38.72 &  30.29 &          &          &          &           &  28.50   &  31.1               &                   \\
$\Gamma_{c}$    &  37.39 &  30.00 &          &          &          &           &  28.10   &  28.6               &                   \\
$\Gamma_{c}$    &  29.67 &  21.22 &          &          &          &           &          &                     &                   \\
$\Gamma'_{2c}$  &  24.71 &  16.19 &  14.54   &  14.41   &          &           &          &                     &  15.3$\pm$0.5$^g$ \\
$\Gamma_{15c}$  &  14.68 &   7.95 &   7.63   &   7.38   &   7.44   &    7.43   &          &                     &   7.3$^h$         \\
$\Gamma'_{25v}$ &   0.00 &   0.00 &   0.00   &   0.00   &   0.00   &    0.00   &   0.0    &                     &   0.00            \\
$\Gamma_{1v}$   & -29.08 & -23.94 & -22.88   & -22.09   &          &           & -22.0    &  -24.0 $\pm$ 0.5    & -23.0$\pm$0.2$^i$ \\   
\hline                                                                                                                                 
$L'_{2c}$       &  26.97 &  18.58 &  18.14   &  17.47   &          &           &          &                     &  20$\pm$1.5$^j$   \\      
$L_{15c}$       &  18.30 &  11.15 &  10.23   &          &  10.44   &   10.38   &          &                     &                   \\
$L'_{v}$        &  -3.78 &  -3.13 &  -2.98   &          &  -3.01   &   -2.94   &          &                     &                   \\
$L'_{2v}$       & -17.94 & -15.15 & -14.27   & -13.97   &          &           &          &                     & -12.8$\pm$0.3$^j$ \\
$L'_{3v}$       & -21.89 & -18.02 & -16.95   & -16.58   &          &           &          &                     & -15.2$\pm$0.3$^j$ \\
\hline                                                                                                                             
$X_{c}$         &  50.48 &  42.75 &          &          &          &           &  39.90   &  57.2               &                   \\
$X_{c}$         &  44.09 &  36.05 &          &          &          &           &  32.77   &  36.2               &                   \\
$X_{c}$         &  36.13 &  26.95 &          &          &          &           &  24.46   &  23.9               &                   \\
$X_{c}$         &  29.83 &  20.85 &  19.50   &          &          &           &          &  23.9               &                   \\
$X_{c}$         &  13.31 &   6.30 &   6.30   &          &   6.23   &    6.26   &          &                     &                   \\
$X'_{4v}$       &  -8.38 &  -7.11 &  -6.69   &          &  -6.72   &   -6.58   &  -6.74   &  -6.97 $\pm$ 0.075  &                   \\   
$X'_{1v}$       & -17.78 & -14.68 & -13.80   &          &          &           & -13.42   & -14.01 $\pm$ 0.075  &                   \\
\hline                                                                                                                              
$K_{c}$         &  12.29 &   7.49 &          &          &   7.22   &           &          &                     &                   \\
$K'_{1v}$       &  -9.29 &  -6.05 &          &          &  -5.74   &           &          &                     &                   \\
\hline                                                                                                                                
$W_{c}$         &  20.50 &  13.02 &          &          &  12.43   &           &          &                     &                   \\
$W_{v}$         & -10.84 &  -9.36 &          &          &  -8.85   &           &          &                     &                   \\
\hline                                                                                                                                
$E_g$           &  12.63 &   5.75 &   5.67   &   5.75   &   5.63   &           &          &                     &   5.48$^j$        \\
\end{tabular}
\begin{tablenotes}
\item[a]$^a$ This work GO
\item[b]$^b$ Ref. [\onlinecite{Rohlfing93}] GO  \\
\item[c]$^c$ Ref. [\onlinecite{Lofas11}] PW     \\
\item[d]$^d$ Ref. [\onlinecite{Gao15}] PW       \\
\item[e]$^e$ Ref. [\onlinecite{Nabok16}] FPLAPW \\
\item[f]$^f$ Ref. [\onlinecite{Edmonds13}] PW \goLDA and CIS photoemission\\
\item[g]$^g$ Ref. [\onlinecite{Himpsel80}]      \\
\item[h]$^h$ Ref. [\onlinecite{Roberts67}]      \\
\item[i]$^i$ Ref. [\onlinecite{Jimenez97}]      \\
\item[j]$^j$ Ref. [\onlinecite{Hellwege82}]     \\
\end{tablenotes}
\end{ruledtabular}
\label{tab3}
\end{table*}

The GO HF band structure for diamond in Table \ref{tab3} and Fig. \ref{fig6} shows a valence band width of 29.07 eV, in good agreement 
with previously reported HF values of 28.67 eV \cite{Barnard02} or 29.43 eV \cite{Stoyanova14}. When \gowo self-energy corrections are made,
the valence band width reduces to 23.94 eV, in good agreement with the range of measured values
from photoemission \cite{Jimenez97,Yokoya06,Edmonds13}. Direct and indirect HF band gaps, 14.71 and 12.62 eV, 
are in good agreement with previously reported values of 14.7 and 12.6 eV \cite{Shimazaki08}. The \goHF direct gap is 
7.95 eV, 0.6 eV greater than the experimental value of Roberts and Walker \cite{Roberts67}, namely 7.3 eV. A previous GO \goLDA calculation 
\cite{Rohlfing93} predicted a direct gap about 0.3 eV greater than the experimental value. \goLDA valence band widths are 
1-2 eV less than \goHF values. The \goHF indirect gap in diamond is 5.75 eV, 0.3 eV greater than the experimental value of 5.48 eV 
\cite{Hellwege82}.

Edmonds and coworkers \cite{Edmonds13} reported a constant initial state (CIS) photoemission study of the (100) surface of
H-terminated diamond and gave precise values of valence band energies at the $X$ point. They found the lower and upper VB
at -14.01 and -6.97 eV (Table \ref{tab3}), which compares to -14.68 and -7.11 eV from \goHF and -13.42 and -6.74 eV
from \goLDA with a PW basis. Table \ref{tab3} also compares positions of higher CB at the $\Gamma$ and $X$ points and these
are in reasonable agreement with \goHF and \goLDA calculations, although it should be noted that experimental error bars are
several eV wide \cite{Edmonds13}.

\subsection{Silicon}

\begin{table}[ht!]
\caption{HF, \gowop @HF and \gowop @LDA conduction and valence band energies for Si at high symmetry points relative to 
the VBM at the $\Gamma$ point.}
\begin{ruledtabular}
%See Johannsson_90 for higher bands
\begin{tabular}{lcccccc}
State           & HF     &\gowop@HF &\gowop@LDA$^a$&\gowop@LDA$^b$& Expt$^c$                 \\ % GW@TDHF      & HF Duchemin
\hline                                                                                     
$\Gamma'_{2c}$  &  10.63 &   5.19   &   4.27       &  4.08        &  4.2                     \\ %  &   4.33    &    
$\Gamma_{15c}$  &   8.41 &   3.53   &   3.30       &  3.35        &  3.35$^d$                \\ %  &   2.86    &  9.12   
$\Gamma'_{25v}$ &   0.00 &   0.00   &   0.00       &  0.00        &                          \\ %  &   0.00    &   
$\Gamma_{1v}$   & -17.00 & -13.03   &              & 12.04        & 12.5$\pm$0.6             \\ %  & -11.53    &   
\hline                                                                                                           
$L'_{2c}$       &   9.66 &   4.54   &   4.11       &  4.24        & 4.15$\pm$0.1             \\ %  &   3.87    &   
$L_{15c}$       &   7.81 &   2.84   &   2.30       &  2.27        & 2.1$^e$,2.4$\pm$0.15$^f$ \\ %  &   2.47    &  7.92 
$L'_{3v}$       &  -1.94 &  -1.51   &  -1.19       & -1.27        & 1.2$\pm$0.2,1.5$^g$      \\ %  &  -1.62    &   
$L'_{2v}$       &  -9.67 &  -7.73   &              & -7.18        &  6.7$\pm$0.2             \\ %  &  -6.87    &   
$L'_{1v}$       & -14.01 & -10.87   &              & -9.79        &  9.3$\pm$0.4             \\ %  &  -9.55    &   
\hline                                                                                                           
$X_{1c}$        &   6.37 &   1.47   &              &  1.44        &                          \\ %  &   1.14    &  6.62 
$X'_{4v}$       &  -4.29 &  -3.48   &              & -2.99        & 3.3$\pm$0.2$^h$          \\ %  &  -3.44    &   
$X'_{2v}$       & -11.38 &  -8.96   &              &              &                          \\ %  &  -7.91    &   
\hline                                                                                                           
$K_{1c}$        &   6.59 &   1.59   &              &              &                          \\ %  &   1.66    &   
$K'_{4v}$       &  -4.03 &  -3.36   &              &              &                          \\ %  &  -2.96    &   
\hline                                                                                                           
$W_{1c}$        &  10.48 &   5.13   &              &              &                          \\ %  &   4.62    &   
$W'_{4v}$       &  -5.23 &  -4.43   &              &              &                          \\ %  &  -4.12    &   
\hline                                                                                                           
$E_g$           &   6.21 &   1.38   &   1.24       & 1.29         &  1.17                    \\ %  &   1.09    &  6.43   
\end{tabular}
\begin{tablenotes}
\item[a] $^a$Ref. [\onlinecite{Godby88}]
\item[b] $^b$ Ref. [\onlinecite{Hybertsen86}] \\
\item[c] $^c$ Ref. [\onlinecite{Hellwege82}] except where noted\\
\item[d] $^d$ Ref. [\onlinecite{Lautenschlager87}] \\
\item[e] $^e$ Ref. [\onlinecite{Hulthen76}] \\
\item[f] $^f$ Ref. [\onlinecite{Straub85}] \\
\item[g] $^g$ Ref. [\onlinecite{Himpsel81}] \\
\item[h] $^h$ Ref. [\onlinecite{Wachs85}] \\
\end{tablenotes}
\end{ruledtabular}
\label{tab4}
\end{table}

\begin{table}[ht!]
\caption{Critical point transition energies for Si from \gowop @HF, \gowop @LDA calculations and ellipsometry.}
\begin{ruledtabular}
\begin{tabular}{lcccccc}
Transition                           &             &\gowop@HF &\gowop@LDA$^a$&\gowop@LDA$^b$& Expt$^c$  \\
\hline                                                                                                           
$\Gamma_{25v}^{'} \to \Gamma_{15c}$  &   $E_0^{'}$ &   3.53   &   3.30       &  3.35        &  3.35      \\
$L_{3v}^{'} \to L_{1c}$              &   $E_1$     &   4.35   &   3.54       &  3.49        &  3.46      \\
$L_{3v}^{'} \to L_{3c}$              &   $E_1^{'}$ &   6.05   &   5.30       &  5.51        &  5.4       \\
$X_{4v}     \to X_{1c}$              &   $E_2$     &   4.95   &   4.43       &   -          &  4.32      \\
\end{tabular}
\begin{tablenotes}
\item[a] $^a$Ref. [\onlinecite{Godby88}]
\item[b] $^b$Ref. [\onlinecite{Hybertsen86}] \\
\item[c] $^c$Ref. [\onlinecite{Lautenschlager87}] \\
\end{tablenotes}
\end{ruledtabular}
\label{tab5}
\end{table}

HF and \goHF band structures of Si are shown in Fig. \ref{fig6} and energies of valence and conduction band states
at high symmetry points are compared to experimental values and previous \gowop @LDA calculations in Table \ref{tab4}. The HF valence 
bandwidth of Si in this work is 17.00 eV, considerably larger than the experimental value of 12.5 $\pm$ 0.6 eV \cite{Hellwege82}. 
The bottom of the VB has 3$s$ character and the top has 3$p$ character; the \goHF self-energy for the bottom of the band is 
larger than the top, resulting in a narrowing of the valence bandwidth to 13.03 eV. The predicted VB width lies within
the experimental uncertainty range and compares to a \goLDA value of 12.04 eV \cite{Hybertsen86} (Table \ref{tab4}). 
Duchemin and Gygi \cite{Duchemin10} reported the HF band structure for Si using a PW basis.
They obtained conduction band energies of 9.12, 7.92 and 6.62 eV at the $\Gamma$, L and X points of the Brillouin zone,
and an indirect gap of 6.43 eV which compares to 8.41, 7.81 and 6.37 eV in this work and 6.21 eV for the indirect gap (Table \ref{tab4})
in this work using an all-electron GO basis.

The \gowop @DFT band structure of Si is well established \cite{Hybertsen86,Godby88,Rohlfing95}. These data and photoemission and optical 
transition measurements allow a detailed comparison of the \goHF calculations. A comparison of critical point
transition energies from theory and experiment is given in Table \ref{tab5}. \goLDA calculations agree with ellipsometry values 
to within 0.1 eV while \goHF values overestimate transition energies by between 0.2 eV (E$_0$) and 0.9 eV (E$_1$).

\subsection{Rock Salt MgO}

\begin{figure}[htp]
\includegraphics[width=6cm,angle=0]{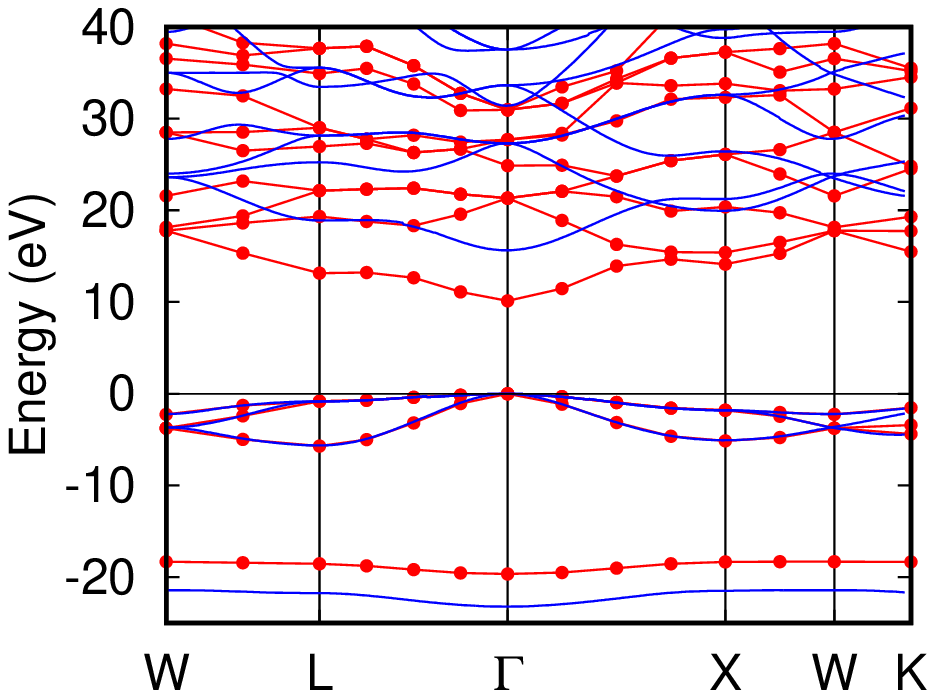}
\caption{Band structures of rock salt MgO from HF and \gowo calculations. HF band structures are shown as solid blue lines. \gowo 
self-energies were obtained in an 8x8x8 MP mesh at points shown.}
\label{fig7}
\end{figure}

The \goHF band structure of MgO calculated is shown in Fig. \ref{fig7} and conduction and 
valence band energies are compared to results of a previous \goLDA calculation and the experimental gap in Table \ref{tab6}.
R{\"o}ssler and Walker \cite{Roessler67} reported an exciton binding energy of 0.08 eV with an 
exciton dissociation limit of 7.77 eV from optical reflectance measurements and Walker and coworkers reported exciton fine structure 
between 7.67 and 7.83 eV \cite{Whited73}. Sch{\"o}nberger and Aryasetiawan \cite{Schonberger95} 
reported a \goLDA band gap of 7.7 eV using an LMTO-ASA method. More recently, band gaps of 7.90 eV \cite{Gao16} 
(PW/LDA), 7.63 eV \cite{Nabok16} (FLAPW/LDA), 7.32 eV \cite{Ren21} (numerical atomic orbitals/PBE), 8.53 eV \cite{Begum21} 
(FLAPW/HSE06) and 9.97 eV \cite{SalasIllanes22} quasiparticle self-consistent $GW$ (QSGW) have been reported. The \goHF gap value from 
this work is 9.13 eV. Antonius and coworkers \cite{Antonius15} reported zero point renormalization reduction in the band gaps of 
diamond and MgO by over 0.3 eV in diamond and nearly 0.3 eV in MgO. First principles calculations omit screening by polar phonons, 
which is, of course, affects experiental measurements. 

The value of 9.13 eV for the band gap of MgO from \goHF is therefore about 1.1 eV above the experimental gap of 7.77 eV, taking these
effects into account. It is 0.6 eV higher than the FLAPW/HSE06 value \cite{Begum21} mentioned above.
The HF and \goHF band structures for MgO in Fig. \ref{fig7} show that the O 2$p$ valence band widths are not strongly affected
by self energy corrections, while the O 2$s$ band shifts upwards relative to the VBM by 4.0 eV.

\begin{table}[ht!]
\caption{HF, \goHF  and \goLDA lowest conduction band and highest valence band energies for MgO relative to the VBM at 
the $\Gamma$ point.}
\begin{ruledtabular}
\begin{tabular}{lccccccc}
State           & HF     &\goHFp$^a$&\goLDA$^b$ & Expt$^c$ \\   %& \GWTDHF 
\hline                                                                      
$\Gamma_{c}$    &  15.97 &   9.13  &     7.32  & 7.77(8.1)$^d$ \\   %&    9.21 
$\Gamma_{v}$    &   0.00 &   0.00  &     0.00  &          \\   %&    0.00 
\hline                                                                   
$L_{c}$         &  18.78 &  11.74  &    10.46  &          \\   %&   11.47 
$L_{v}$         &  -0.95 &  -0.98  &    -0.77  &          \\   %&   -0.79 
\hline                                                                   
$X_{c}$         &  19.83 &  12.75  &    11.43  &          \\   %&   12.74 
$X_{v}$         &  -1.89 &  -1.98  &    -1.56  &          \\   %&   -1.73 
\end{tabular}                                                     
\begin{tablenotes}
\item[a]$^a$ This work GO
\item[b]$^b$ Ref. [\onlinecite{Nabok16}] \\                               
\item[c]$^c$ Ref. [\onlinecite{Whited73}] \\                              
\item[d]$^d$ Estimate without electron-phonon interaction Ref. [\onlinecite{Antonius15}] \\                              
\end{tablenotes}
\end{ruledtabular}                                                
\label{tab6}                                                      
\end{table}                                                       
                                                                  
\subsection{Anatase and Rutile TiO$_2$}                           
                                                                  
\begin{figure}[htp]
\includegraphics[width=6cm,angle=0]{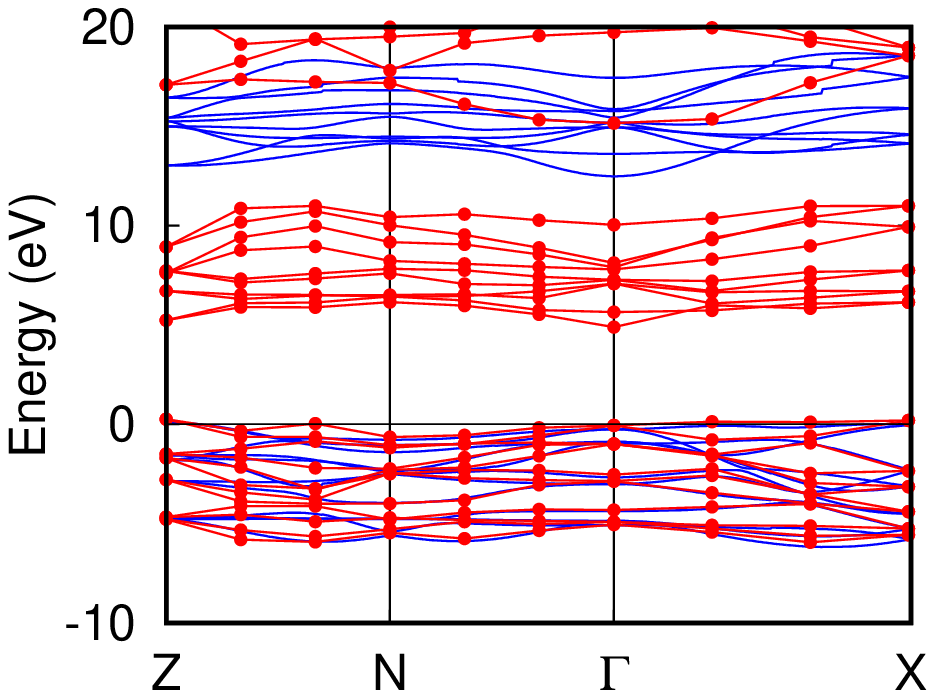}
\caption{Band structures of anatase TiO$_2$ from HF and \goHF calculations. The HF band structure is shown as solid blue lines. \goHF
self-energies were obtained in a 6x6x6 MP mesh at points shown.}
\label{fig8}
\end{figure}

Angle resolved photoelectron spectroscopy (ARPES) measurements on anatase TiO$_2$ at 20 K show that the VBM occurs at the $X$ point of 
the BZ and that it lies approximately 0.5 eV above the highest band at the $\Gamma$ point \cite{Baldini18}. The same work found a 
QP band gap at $\Gamma$ of 3.97 eV \cite{Baldini18} using $n$-type samples with O vacancies. HF and \goHF band energies for anatase 
are compared to \goLDAp, \goPBE values and ARPES data in Table \ref{tab7} and the band structures are shown in Fig. \ref{fig8}.

Kang and Hybertsen \cite{Kang10} found the VBM close to the $X$ point along the $\Gamma-X$ direction, the CBM at $\Gamma$ and
an indirect gap of 3.56 eV and direct gaps at $\Gamma$ and $X$ of 4.14 and 4.95 eV, respectively. Baldini and coworkers
found the VBM at $X$ and the CBM at $\Gamma$ \cite{Baldini18} with an indirect gap of 3.46 eV, in good agreement with ARPES data
in the same work. 

\begin{table}[ht!]
\caption{HF, \goHFp, \goLDA and \goPBE lowest conduction band and highest valence band energies for anatase TiO$_2$ relative to the VBM at 
the $X$ point.}
\begin{ruledtabular}
\begin{tabular}{lccccc}
State         & HF    &\goHFp$^a$ &\goLDA$^b$ & \goPBEp$^c$ & Expt$^c$ \\
\hline                                 
$\Gamma_{c}$  & 12.80 &   4.83    &  3.56     &  3.46       &  3.47 \\
$\Gamma_{v}$  & -0.27 &  -0.29    & -0.58     & -0.46       & -0.50 \\
\hline                                                        
$X_{c}$       & 14.30 &   6.02    &  4.89     &  -          &  -    \\
$X_{v}$       &  0.00 &   0.00    & -0.06     &  0.00       &  0.00 \\
%$\Gamma_{c}$  & 12.70 &   5.00    &  3.56     &  3.46       &  3.47 \\
%$\Gamma_{v}$  & -0.54 &  -0.56    & -0.58     & -0.46       & -0.50 \\
%\hline                                                        
%$X_{c}$       & 14.32 &   6.29    &  4.89     &  -          &  -    \\
%$X_{v}$       &  0.00 &   0.00    & -0.06     &  0.00       &  0.00 \\
\end{tabular}
\begin{tablenotes}
\item[a]$^a$ This work GO
\item[b]$^b$ Ref. [\onlinecite{Kang10}] \\
\item[c]$^b$ Ref. [\onlinecite{Baldini18}] \\
\end{tablenotes}
\end{ruledtabular}
\label{tab7}
\end{table}

\begin{figure}[htp]
\includegraphics[width=6cm,angle=0]{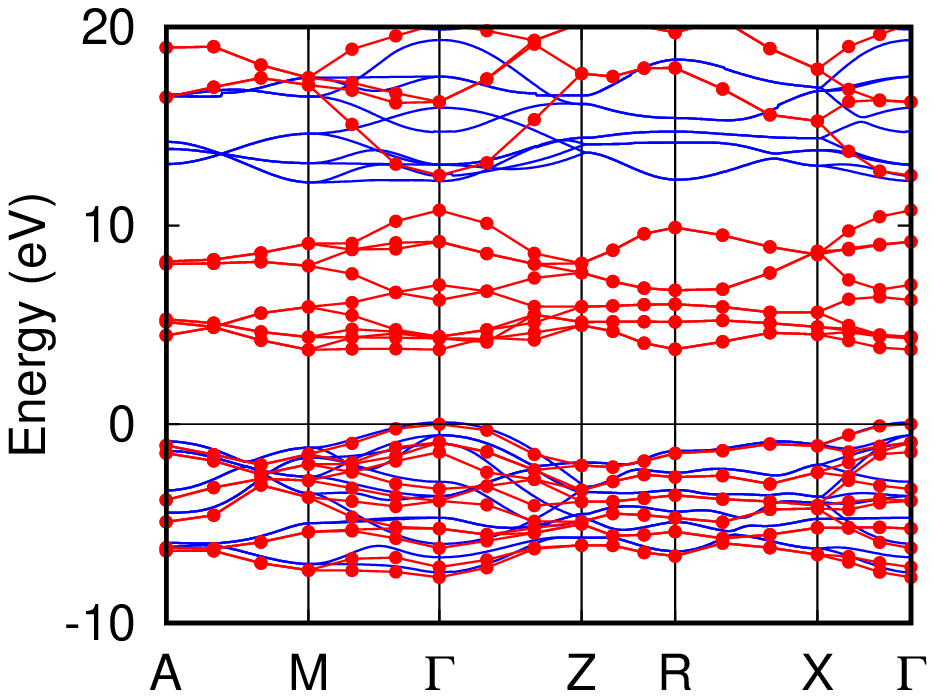}
\caption{Band structures of rutile TiO$_2$ from HF and \gowo calculations. HF band structures are shown as solid blue lines. \gowo 
self-energies were obtained in a 6x6x6 MP mesh at points shown.}
\label{fig9}
\end{figure}

Low temperature, high resolution absorption measurements on rutile TiO$_2$ show the presence of a dark exciton at 3.03 eV,
which has been used to infer a direct QP gap at $\Gamma$ for rutile \cite{Pascual77}.
Kang and Hybertsen \cite{Kang10} found an indirect gap of 3.34 eV in rutile, with the VBM at the $R$ point. However, the highest VB
at $\Gamma$ is just 40 meV below that at the $R$ point. More recent calculations by Baldini and coworkers \cite{Baldini17} 
found a direct gap of 3.34 eV. Similar values for the rutile QP gap in the range 3.30 \cite{Zhang15} to 3.59 eV \cite{Chiodo10}
have also been reported. 
The direct gap at $\Gamma$ from \goHF is 3.85 eV and is less than the indirect gap values of 3.88 and 3.94 eV at the $M$ and $R$ points, 
respectively. The \goHF band structure is shown in Fig. \ref{fig9} and band edge energies are compared to \goLDA values in Table
\ref{tab8}.

The direct gap at $\Gamma$ from \goHF for anatase is 5.12 eV, larger than the experimental value of 3.97 eV \cite{Baldini18}.
Thus the gaps in anatase and rutile TiO$_2$ exceed experimental or \goLDA values by 0.5 eV or more.
Band ranges used in the \goHF self-energy calculation for both anatase and rutile TiO$_2$ (Table \ref{tab2}) were 12 valence bands 
and 20 conduction bands and a cutoff of 30 eV. This band range requires diagonalization of RPA matrices of dimension 51840 at the unique 
$\bQ$ points in the 6x6x6 MP mesh (30 for anatase and 40 for rutile). This cutoff energy is much lower than the smaller diamond, Si and MgO 
unit cells owing to the larger numbers of bands at low energies. A higher cutoff may therefore increase the self-energy magnitudes
and reduce the valence-conduction band gap.
%Wang and coworkers \cite{Wang24} with dense k-meshes, place the  direct gap for anatase TiO$_2$ at 3.92 eV \cite{Baldini17}.
                                                                  
%The QP gap in \goHF is 4.41 eV at $\Gamma$, 
%gap renormalization \cite{Ohad23}
%rutile gap \cite{Tezuka94,Hardman94,Rangan10} rutile TiO2 is direct gap \cite{Pascual77,Pascual78}

\begin{table}[ht!]
\caption{HF, \goHF  and \goPWLDA lowest conduction band and highest valence band energies for rutile TiO$_2$ relative to the VBM at 
the $\Gamma$ point.}
\begin{ruledtabular}
\begin{tabular}{lcccc}
State           & HF    &\goHFp$^a$ & \goLDA$^b$ \\ %& \GWTDHF 
\hline                                              
$\Gamma_{c}$    & 12.41 &   3.85   &  3.38      \\ %&    2.60 
$\Gamma_{v}$    &  0.00 &   0.00   &  0.00      \\ %&    0.00 
\hline                                             %          
$R_{c}$         & 12.54 &   4.00   &  3.34      \\ %&    2.68 
$R_{v}$         & -1.38 &  -1.49   & -1.12      \\ %&   -1.44 
\hline                                             %          
$M_{c}$         & 12.34 &   3.89   &  3.40      \\ %&    2.64 
$M_{v}$         & -1.16 &  -1.48   & -1.15      \\ %&  - 1.54 
%$\Gamma_{c}$    & 12.33 &   3.85   &  3.38      \\ %&    2.60 
%$\Gamma_{v}$    &  0.00 &   0.00   &  0.00      \\ %&    0.00 
%\hline                                             %          
%$R_{c}$         & 12.38 &   3.94   &  3.34      \\ %&    2.68 
%$R_{v}$         & -1.34 &  -1.44   & -1.12      \\ %&   -1.44 
%\hline                                             %          
%$M_{c}$         & 12.24 &   3.88   &  3.40      \\ %&    2.64 
%$M_{v}$         & -1.20 &  -1.50   & -1.15      \\ %&  - 1.54 
\end{tabular}
\begin{tablenotes}
\item[a]$^a$ This work GO
\item[b]$^b$ Ref. [\onlinecite{Kang10}] \\
\end{tablenotes}
\end{ruledtabular}
\label{tab8}
\end{table}

\section{Dielectric Functions from \goHFp/BSE\label{dielectric}}

We present dielectric functions from \goHFp/BSE-TDA calculations in this Section. In the previous Section, nxnxn MP meshes with n = 8 or 
n = 6 were used for self-energy corrections to band structures. Points in the BZ in the Exciton code must be integers modulo n. 
BZ high symmetry points for the lattices used contain reciprocal lattice vector fractions such as $\mathbf{G}$/4 or $\mathbf{G}$/3.
Consequently, to construct straight lines in reciprocal space between high symmetry points (apart from the $\Gamma$ point), MP mesh dimensions
for cubic diamond, Si or MgO must be multiples of 4 and those for anatase and rutile TiO$_2$ must be multiples of 3. For BSE-TDA calculations 
of dielectric functions this restriction is lifted as all k-points are used.

BSE-TDA calculations reported in this Section used 10x10x10 or 11x11x11 MP meshes centered on the $\Gamma$ point.
Symmetric meshes permit symmetry to be used to reduce the number of $\bQ$ points at which the RPA-TDA Hamiltonian (Eq. \ref{eqn22}) 
must be diagonalized in order to construct the dressed polarizability on which the screened interaction depends.
Symmetry equivalence of k-points means limited sampling of the BZ, however. 
Dielectric functions presented demonstrate that the method produces dielectric spectra which agree with experiment reasonably well. 
Methods of improving convergence of the approach are discussed later in Section \ref{discussion}.

A study of oscillator strengths in molecules \cite{Caricato11}, which compared oscillator strengths from 
a range of DFT exchange-correlation functionals using Kohn-Sham orbitals and RPA and CIS methods using HF orbitals
to EOM-CCSD, found that oscillator strengths from RPA and CIS methods typically exceeded those from the reference method while
those from DFT underestimate them. The difference depended on the bonding type - for alkenes, CIS overestimates them by a factor of 1.45
on average, while LDA underestimates by a factor of 0.86. For azabenzenes, CIS overestimates them by a factor of 2.22 on average,
while LDA is almost in agreement with EOM-CCSD (0.99). The CIS method is BSE-TDA with no screening of the electron-hole interaction 
and is the method most similar to BSE-TDA of the methods used in Ref. [\onlinecite{Caricato11}]. We find similar overestimates of oscillator 
strength in the crystal systems studied here using HF wave functions. For diamond and Si there is reasonable agreement with experimental 
oscillator strengths, while for the oxides chosen for study, there is significant overestimation of oscillator strengths.

\subsection{Diamond}

\begin{figure}[htp]
\includegraphics[width=4.25cm,angle=0]{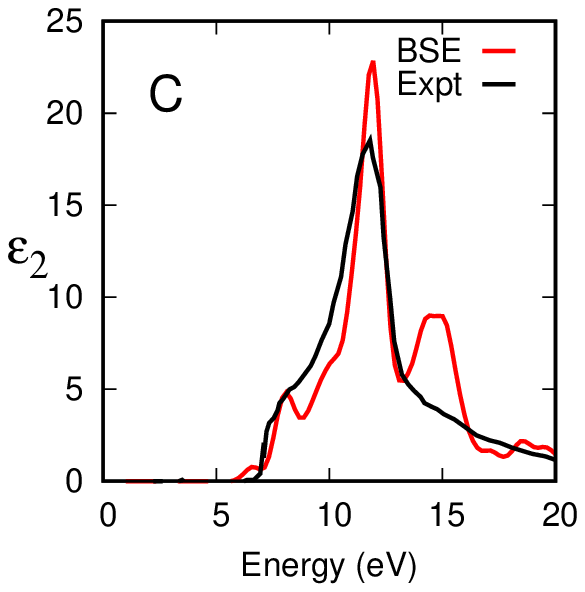}
\includegraphics[width=4.25cm,angle=0]{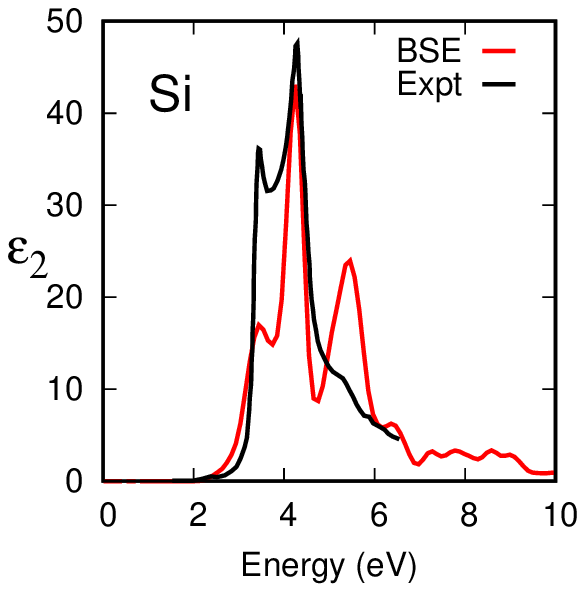}
\caption{Dielectric functions of diamond (left panel) and Si (right panel) from BSE calculations which used \goHF
self-energies as input to the BSE-TDA calculation. Experimental data redrawn from Ref. [\onlinecite{OpticalHandbook}] is shown in 
solid black lines.}
\label{fig10}
\end{figure}

The dielectric function of diamond from \goHFp/BSE is compared to experimental data in Fig. \ref{fig10}(a). It was obtained by averaging 
spectra with n = 10 and n = 11 to improve BZ sampling and by smoothing with a spline fit. The self-energy calculation used four valence 
and 10 conduction bands in the \goHF self-energy (Eq. \ref{eqn34}) plus a further 20 conduction bands in the second order self-energy 
(Eq. \ref{eqn33}). The BSE-TDA calculation used the same bands as the \goHF self-energy. The experimental lineshape and intensity 
is recovered well except for overemphasis of a peak around 15 eV. Peak positions in the BSE-TDA and experiment are in good agreement (11.8
eV in both). Compared to previous calculations of the dielectric function of diamond in a GO basis \cite{Patterson20a}, which used
the HF band structure and a TDHF Hamiltonian with a uniformly reduced electron-hole attraction, the width and position of the absorption 
peak from \goHFp/BSE is in much better agreement with experiment then before.

\subsection{Silicon}

The dielectric function of Si from \goHFp/BSE is compared to experimental data in Fig. \ref{fig10}(b). It was obtained from a single
calculation with n = 10. The self-energy calculation used four valence and 14 conduction bands in the \goHF self-energy (Eq. \ref{eqn34})
plus a further 32 conduction bands in the second order self-energy (Eq. \ref{eqn33}). The BSE-TDA calculation used the same bands as the
\goHF self-energy. Peak positions in the dielectric function are in good agreement with experimental values (3.45, 4.24 and 5.43 eV 
versus 3.44, 4.29 and 5.30 eV in experiment), however the intensity of the first peak (which is enhanced by electron-hole attraction) 
is about a factor of two too low and the weak shoulder at 5.30 eV in experiment is a strong peak in the calculation.

\subsection{Rock Salt MgO}

\begin{figure}[htp]
\includegraphics[width=4.5cm,angle=0]{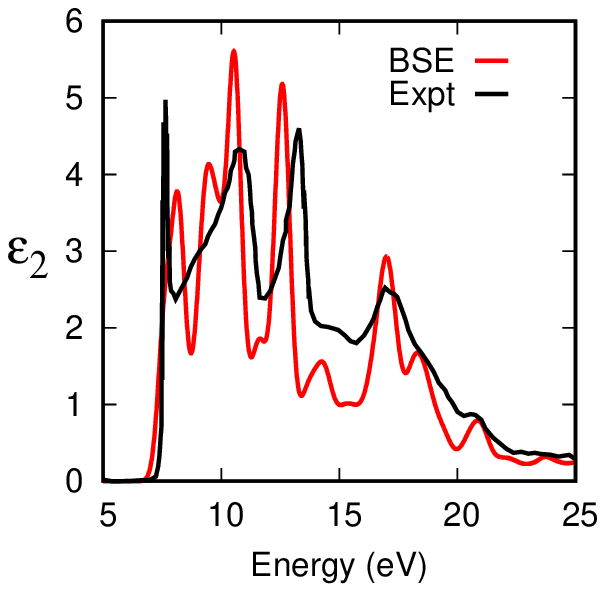}
\caption{Dielectric functions of MgO from BSE-TDA calculations which used \goHF self-energies as input to the BSE-TDA Hamiltonian. 
Experimental data redrawn from Ref. [\onlinecite{Roessler67}] is shown in solid black lines.}
\label{fig11}
\end{figure}

The BSE-TDA dielectric function of MgO is shown in Fig. \ref{fig11} along with the experimental dielectric function from Ref.
[\onlinecite{Roessler67}]. A single n = 10 calculation with four valence bands, 17 conduction bands in the the \goHF self-energy 
(Eq. \ref{eqn34}) plus a further 32 conduction bands in the second order self-energy (Eq. \ref{eqn33}).

Peak positions in the BSE-TDA dielectric function (8.09, 10.50, 12.64 and 16.93 eV) are in reasonable agreement with experimental values 
(7.6, 10.8, 13.3 and 16.9 eV). Unlike diamond and silicon, peak intensities from BSE-TDA are greater than experiment. Similar 
overestimation in the dielectric function intensities are also found in anatase and rutile TiO$_2$. The tetrahedral semiconductors
considered here have delocalized, nearly free electron wave functions while the polar, ionic oxides have more localized occupied HF wave
functions.

The experimental optical gap of MgO is 7.83 eV \cite{Roessler67,Whited73}.
The experimental exciton binding energy is 0.08 eV \cite{Roessler67}. FLAPW/HSE06 calculations \cite{Begum21} predict an exciton
binding energy of 0.44 eV and 0.60 eV using PBEsol and HSE06 DFT starting points in the \gowop/BSE calculations. Here we find
an exciton binding energy of 1 eV. Recent calculations of the renormalization of exciton energies in MgO \cite{Schebek25}
by polar phonons report a redshift of around 0.05 eV for MgO, which is much less than the differences between experimental and 
first principles values.

\subsection{Anatase and Rutile TiO$_2$}

\begin{figure}[htp]
\includegraphics[width=7.5cm,angle=0]{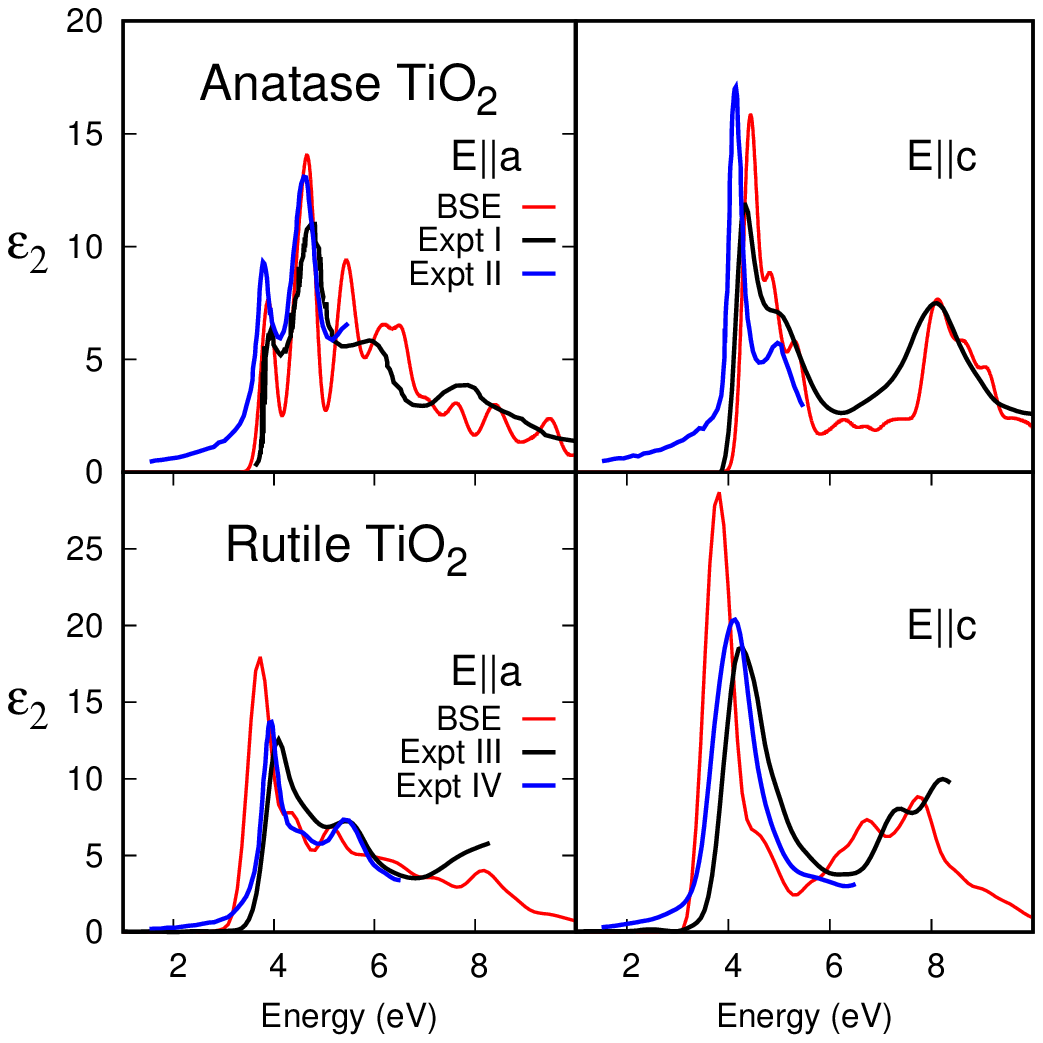}
\caption{Dielectric functions of anatase (upper panel) and rutile (lower panel) TiO$_2$ from \goHFp/BSE-TDA calculations self-energies
as input to the BSE-TDA calculation. BSE spectra were scaled by a factor of 0.6.
Experimental data for anatase TiO$_2$ are redrawn from Refs. [\onlinecite{Hosaka97}] 
(Expt I, optical reflectance, black line) and [\onlinecite{Baldini18}] (Expt II, SE, blue line).
Experimental data for rutile TiO$_2$ are redrawn from Refs. [\onlinecite{Tiwald00}] 
(Expt III, optical reflectance, black line) and [\onlinecite{Baldini17}] (Expt IV, SE, blue line).}
\label{fig12}
\end{figure}

The dielectric functions of anatase and rutile TiO$_2$ from \goHFp/BSE are compared to experimental data in Fig. \ref{fig12}. 
They were obtained using a MP mesh with n = 6. Self-energy and BSE screening calculations for both phases used 12 valence and 20 
conduction bands (Table \ref{tab2}). Hosaka and coworkers obtained the dielectric function of anatase TiO$_2$ from optical reflection data 
\cite{Hosaka97} and Baldini and coworkers \cite{Baldini18} reported the dielectric function of anatase TiO$_2$ from spectroscopic 
ellipsometry (SE) and compared it to results of \goPBEp/BSE-TDA calculations (Table \ref{tab9}). 

\goHFp/BSE-TDA calculations reproduce the experimental peak positions for anatase TiO$_2$ mostly within 0.1 to 0.2 eV. For E$\|$a, 
peaks occur at 4.01 and 4.61 eV (compared to 3.90/3.79 eV and 4.67/4.61 eV from optical reflectance/SE data). For E$\|$c they 
occur at 4.56 and 4.94 eV (compared to 4.27/4.13 eV and 5.01/4.96 eV from experiment). As noted in the introduction to 
Section \ref{dielectric} spectral intensities predicted by \goHFp/BSE for oxides exceed experimental intensities. To facilitate 
comparison of predicted and measured spectra, intensities have been scaled by a factor of 0.6. 
\goPBEp/BSE-TDA calculations in Ref. [\onlinecite{Baldini18}] predict peaks at 3.76 and 4.81 eV for E$\|$a and at 4.28 eV for E$\|$c,
which are close to peak positions in SE experiments at 3.79, 4.61 and 4.13 eV (Table \ref{tab9}).

\noindent
\begin{table}[ht!]
\caption{Peak positions in dielectric functions of anatase and rutile TiO$_2$ from optical reflectivity, spectroscopic ellipsometry
and \goHFp/BSE-TDA and \goPBEp/BSE-TDA calculations.}
\begin{ruledtabular}
\begin{tabular}{cccccc} 
   Field    & Opt. Refl.$^a$ & Spec. Ellips.$^b$ & BSE-TDA$^b$& BSE-TDA$^c$  \\
\hline                                                        
            &                & Anatase           &            &              \\  %
\hline                                                                            
  E$\|$a    &   3.90         &   3.79            & 3.76       &  4.01        \\  %
            &   4.67         &   4.61            & 4.81       &  4.78        \\  %
  E$\|$c    &   4.27         &   4.13            & 4.28       &  4.56        \\  %
            &   5.01         &   4.96            &            &  4.94        \\  %
            &   8.08         &                   &            &  8.24        \\  %
\hline                                                                            
            &                & Rutile            &            &              \\  %
\hline                                                                            
  E$\|$a    &   4.05         &   3.93            & 3.99       &  3.71        \\  %
            &                &   4.51            & 4.57       &  4.32        \\  %
            &   5.43         &   5.42            & 5.37       &  5.12        \\  %
  E$\|$c    &   4.20         &   4.15            & 4.24       &  3.80        \\  %
\end{tabular}
\begin{tablenotes}
\item[a]$^a$ Ref. [\onlinecite{Hosaka97}]
\item[b]$^b$ Ref. [\onlinecite{Baldini18}]
\item[c]$^c$ This work GO
\end{tablenotes}
\end{ruledtabular}
\label{tab9}
\end{table}

%The dielectric function of rutile TiO$_2$ from \goHFp/BSE is compared to experimental data in the lower panels of Fig. \ref{fig12}. 
Tiwald and Schubert obtained the dielectric function of rutile TiO$_2$ from optical reflectivity \cite{Tiwald00} and Baldini and
coworkers obtained it from spectroscopic ellipsometry \cite{Baldini17}. For E$\|$a, SE shows a peak at 3.93 eV, a shoulder at 
4.51 eV and a further peak at 5.42 eV (Table \ref{tab9}). Optical reflectivity shows a peak at 4.05 eV and a second peak at 5.43 eV. 
\goPBE/BSE-TDA calculations by Baldini and coworkers \cite{Baldini17} find peaks to match each of these features at 3.99, 4.57 and 5.37 eV.

\goHFp/BSE-TDA calculations reproduce the experimental peaks but are shifted to lower energy by 0.2 to 0.4 eV. For E$\|$a peaks 
occur at 3.71, 4.32 eV and 5.12 eV (0.22, 0.19 and 0.30 eV below the SE values) and for E$\|$c the main peak occurs at 3.80 eV 
(0.35 eV below the SE value). The cause of this shift is believed to be the small $\bQ$ approximation to the screened electron-hole 
attraction described in Section \ref{smallq}. Bare electron-hole attraction terms in the small $\bQ$ limit for anatase and rutile TiO$_2$  
with an n = 6 MP mesh are 1.56 and 0.94 eV, respectively. When screened these become 1.03 and 0.65 eV, corresponding to long wavelength
dielectric functions of 1.45 and 1.50, well below a value of 5 which might be expected in a moderate gap energy oxide. If a value of
5 were used instead, the screened small $\bQ$ limit electron-hole attraction terms would become 0.31 eV and 0.19 eV. The larger 
value for rutile TiO$_2$ may explain the tendency for peaks to be shifted to lower energy in that case. This point is discussed further in Section \ref{discussion}.

\section{Discussion\label{discussion}}

Most \gwbse calculations for periodic, solid materials are performed using codes which employ PW basis sets. These codes have 
been referenced throughout this work and our GO basis results have been compared to them. GO basis $GW$ and BSE
we have presented a \gwbse method 

Early \goLDA calculations by Rohlfing and coworkers \cite{Rohlfing93} used a model dielectric function to treat the band structures
of five tetrahedral semiconductors. Later these authors adopted the dielectric band structure plasmon pole model \cite{Baldereschi79}
for the dielectric function in \goLDA calculations applied to bulk Si and the Si(001) surface \cite{Rohlfing95}. Rohlfing and Louie
\cite{Rohlfing00} then applied this \goLDA method to \goLDAp/BSE-TDA calculations applied to inert gas solids, tetrahedral semiconductors
and wide gap insulators. Galami{\'c}-Mulaomerovi{\'c} and Patterson used the dielectric band structure model for the dielectric
function to calculate the band structures \cite{Galamic05} and excitonic absorption spectra \cite{Galamic05a} 
of solid Ne and Ar. In contrast to the current work, that work used the bare coulomb potential in its plane wave representation
and used wave functions and single-particle band energies from the Crystal code \cite{Crystal03}
Here it is represented in the auxiliary CO basis (Eq. \ref{eqn37}). 

Recently we applied the density-fitting approach used here to
perform TDHF calculations with a scaled (rather than screened) electron-hole interaction \cite{Patterson20a} applied to the 
same materials as in this work (except Si). 
Garc{\'i}a-Bl{\'a}zquez and Palacio \cite{GarciaBlazquez25} reported a GO density-fitting approach which used wave functions and 
single-particle band energies from the Crystal code \cite{Crystal23} and using a 'scissors shift' to align hybrid DFT band gaps
with QP gaps.

The current work establishes a density fitting, GO method for \goHFp/BSE-TDA calculations without invoking a plasmon pole approximation. 
To generate the self-energy and screened interaction matrices, diagonalization of a matrix which is of the same size as the final BSE-TDA
matrix, must be diagonalized at each unique $\bQ$ point in the BZ. As the system size grows the number of valence and especially conduction
bands must be increased so that the method is limited by the size of matrices that can be diagonalized in a reasonable amount of time.
However, increased availability of GPU hardware combined with the ELPA code \cite{Karpov25} will increase significantly the sizes of matrices
which can be diagonalized at a reasonable computational cost.

The symmetric MP mesh that is used in the current work results in limited sampling of the BZ, especially affecting optical spectra.
Meshes of the sizes that have been used for BSE-TDA calculations on MgO in this work result in well converged spectra 
\cite{Begum21} when an off-centered mesh is used. Employing symmetry in reciprocal space reduces the amount of time required for
Ewald potential integral generation (Eq. \ref{eqn40}). The Ewald potential has absolutely convergent real and reciprocal space
terms. The wave vector dependence of the real space term is a simple multiplication of an integral by a phase factor containing $\bq$.
On the other hand the wave vector dependence of the reciprocal space term cannot be separated in this way. An approach to
calculating density fitted matrix elements at all k points rather than symmetry unique points would be to use only the real space
part of the Ewald potential for calculation of fitting coefficients. With some well chosen value of $\gamma$ this would limit the
number of three-center integrals to be calculated, real space integrals could be stored in memory and Fourier transformed by multiplying
in the $\bq$-dependent phase factor as needed. Matrices appearing as $\left[ V_{\alpha\beta}^{\bQ-1} \right]^*$ would then become
triple products of matrices of this size with the inverse of the real space part of the Ewald potential on either side and the 
full (uninverted) Coulomb potential in the middle.

The formalism outlined in Section \ref{linear} differs from the usual field theoretic way of introducing the BSE \cite{Strinati88}
and makes clear the connection of the method linear response theory with the additional provision of screening of the Coulomb potential. 
The approach outlined here in which RPA calculations are performed for finite $\bQ$ vectors could easiliy be adapted to permit 
exciton dispersion calculations. 

This formalism was applied recently to ionization energies and optical excitations in molecules \cite{Patterson24,Waide24}.
As in this case, the starting point for the calculations was HFT \cite{Marom12,Bruneval13,Knight16,Bruneval24} rather than DFT. 
In our previous work the mean signed error (MSE) for \goHF in $\pi$-bonded molecules was 0.3 to 0.4 eV, i.e. the ionization energies were 
overestimated by 0.3 to 0.4 eV by \goHFp. For N localized lone pairs states the overestimate was 0.8 to 1.1 eV. Electron affinities were
not calculated, however, these overestimates contribute to overestimation of the QP gap when starting $G_oW_o$ from HFT.
In this work, the overestimate of the \goHF QP gap for diamond is 0.27 eV, for bulk Si it is 0.19 eV, for MgO it is 1.36 eV (1.0 eV
allowing for polar phonon screening). These values seem seem reasonable compared to those obtained for a range of molecules.

\begin{acknowledgements}
This work was supported by Science Foundation Ireland under grant number 19/FFP/6582, SOOMAT. Calculations were performed on the
Boyle cluster, maintained by the Trinity Centre for High Performance Computing and funded by Science Foundation Ireland.
\end{acknowledgements}

\section*{Data Availability}
The data that support the findings of this study are available from the corresponding author upon reasonable request.

\bibliography{paper2}

\end{document}